\documentclass[12pt,preprint]{aastex}
\usepackage{txfonts}

\begin{document}

\title{Effect of Dust on Lyman-alpha Photon Transfer in Optically Thick Halo}

\author{Yang Yang\altaffilmark{1}, Ishani Roy\altaffilmark{2}, Chi-Wang Shu\altaffilmark{1}
and Li-Zhi Fang\altaffilmark{3}}

\altaffiltext{1}{Division of Applied Mathematics, Brown University,
Providence, RI 02912, USA}
\altaffiltext{2}{Computing Laboratory, University of Oxford, Oxford, OX1 3QD, United Kingdom}
\altaffiltext{3}{Department of Physics,
University of Arizona, Tucson, AZ 85721, USA}

\begin{abstract}

We investigate the effects of dust on Ly$\alpha$ photons emergent from an optically thick
medium by solving the integro-differential equation of the radiative transfer of resonant
photons. To solve the differential equations numerically we use the Weighted Essentially Non-oscillatory method (WENO). Although the effects of dust on radiative transfer
is well known, the resonant scattering of Ly$\alpha$ photons makes the problem non-trivial.
For instance, if the medium has the optical depth of dust absorption and scattering to be
$\tau_a \gg 1$, $\tau \gg 1$, and $\tau \gg \tau_a $, the effective absorption optical depth
in a random walk scenario would be equal to $\sqrt{\tau_a(\tau_a+\tau)}$. We show, however,
that for a resonant scattering at frequency $\nu_0$, the effective absorption optical depth
would be even larger than $\tau(\nu_0)$. If the cross section of
dust scattering and absorption is frequency-independent, the double-peaked
structure of the frequency profile given by the resonant scattering is basically dust-independent.
That is, dust causes neither narrowing nor widening of the width of the double peaked profile. One
more result is that the time scales of the Ly$\alpha$ photon transfer in the
optically thick halo are also basically independent of the dust scattering, even when the scattering is
anisotropic.  This is because those time scales are mainly determined by the transfer in the
frequency space, while dust scattering, either
isotropic or anisotropic, does not affect the behavior of the transfer in
the frequency space when the cross section of scattering is
wavelength-independent. This result does not support the
speculation that dust will lead to the smoothing of the brightness distribution of
Ly$\alpha$ photon source with optical thick halo.

\end{abstract}


\newpage

\section{Introduction}

Ly$\alpha$ photons have been widely applied to study the physics of luminous
objects at various epochs of the universe, such as Ly$\alpha$ emitters,
Ly$\alpha$ blob, damped Ly$\alpha$ system, Ly$\alpha$ forest,
fluorescent Ly$\alpha$ emission, star-forming galaxies, quasars at high redshifts as well as
optical afterglow of gamma ray bursts (Haiman et al. 2000; Fardal et al. 2001;
Dijkstra \& Loeb. 2009; Latif et al. 2011).
The resonant scattering of Ly$\alpha$ photons with neutral hydrogen atoms has a
profound effect on the time, space and frequency dependencies of Ly$\alpha$ photons
transfer in an optically thick medium. Ly$\alpha$ photons emergent from an optically
thick medium would carry rich information of photon sources and halo surrounding
the source of the Ly$\alpha$ photon. The profiles of the emission and absorption
of the Ly$\alpha$ radiation
are powerful tools to constrain the mass density, velocity, temperature
and the fraction of neutral hydrogen of the optically thick medium.
Radiation transfer of
Ly$\alpha$ photons in an optically thick medium is fundamentally important.

The radiative transfer of Ly$\alpha$ photons in a medium consisting of
neutral hydrogen atoms has been extensively studied either
analytically or numerically. Yet, there have been relatively few results
which are directly based on
the solutions of the integro-differential equation of the resonant radiative
transfer. Besides the Field solution (Field 1959, Rybicki \& Dell'Antonio 1994), analytical
solutions
with and without dust mostly are based on the Fokker-Planck (P-F)
approximation (Harrington 1973, Neufeld 1990,
Dijkstra et al. 2006). The P-F equation might miss the detailed balance relationship of
resonant scattering (Rybicki 2006), and therefore, the analytical solutions
cannot describe
the formation and evolution of the Wouthuysen-Field (W-F) local thermalization of
the Ly$\alpha$ photon frequency distribution (Wouthuysen 1952, Field 1958), which is
important for the emission and absorption of the hydrogen 21 cm line (e.g. Fang 2009).
The features of the Ly$\alpha$ photon
transfer related to the W-F local thermalization are also missed. An early effort
(Adams et al. 1971) trying to directly solve the integro-differential equation
of the resonant radiative transfer with numerical method. It still is, however, of
a time-independent approximation.

Recently, a state-of-the-art numerical method has been introduced to solve
the integro-differential equation of the radiative transfer with
resonant scattering (Qiu et al. 2006, 2007, 2008, Roy et al. 2009a).
The solver is based on the weighted essentially non-oscillatory (WENO) scheme
(Jiang \& Shu 1996). With the WENO solver, many physical features of the
transfer of Ly$\alpha$
photons in an optically thick medium (Roy et al. 2009b, 2009c, 2010),
which are missed in the Fokker-Planck equation approximations, have been revealed.
For instance, the WENO solution shows that the time scale of the formation of
the W-F local
thermal equilibrium actually is only about a few hundred times of the resonant
scattering. It also shows that the double peaked frequency profile of
the Ly$\alpha$ photon emergent from an optically thick medium does not follow
the time-independent solutions of the
P-F equation. These results directly indicate the needs of re-visiting
problems which
have been studied only via the F-P time-independent approximation.

We will investigate, in this paper, the effects of the dust on the Ly$\alpha$
photons transfer in an optically thick medium. Dust can be produced at epochs of
low and moderate
redshifts, and even at redshift as high as 6 (Stratta et al. 2007).
Absorption and scattering of dust have been used to explain the
observations on Ly$\alpha$
emission and absorption (Hummer \& Kunasz 1980), such as
the escaping
fraction of Ly$\alpha$ photons (Hayes et al. 2010, 2011, Blanc et al. 2010);
the redshift-dependence
of the ratio between Ly$\alpha$ emitters and Lyman Break galaxies
(Verhamme et al. 2008);
and the ``evolution'' of the double-peaked profile (Laursen et al. 2009).

Nevertheless, it is still unclear whether the
time scale of photon escaping
from optically thick halo will be increasing (or decreasing) when the halo is
dusty. It is also
unclear whether the effects of dust absorption can be estimated by the random walk
picture (Hansen \& Oh 2006). As for the dust effect on the double-peaked profile, the current
results given by different studies seem to be contradictory: some claims that
the dust absorption leads to the narrowing of the double-peaked profile (Lauresen et al 2009),
while others result that the width between the two peaks apparently should be increasing
due to the dust absorption (Verhamme et al. 2006). We will focus on these basic problems,
and examine them with the solution of the integro-differential equation of radiative transfer.

This paper is organized in the following way: section 2 presents the theory of the
Ly$\alpha$ photon transfer in an optically thick medium with dust. The equations of
the intensity and flux of resonant photons in a dusty medium are given. We will study
three models of the interaction between dust and photons:  (1)  dust causes only
scattering with photons; (2) dust causes both
scattering and absorption; and (3) dust causes only absorption of photons.
Section 3 gives the solutions of Ly$\alpha$ photons
escaping from an optically thick spherical halos with dust. The dusty effect on
the double-peaked
profile will be studied in Section 4. The discussion and conclusion are given in
Section 5. Some mathematical derivations of the equations and
numerical implementation details are given in the Appendix.

\section{Basic theory}

\subsection{Radiative transfer equation of dusty halo}

We study the transfer of Ly$\alpha$ photons in a spherical halo with
radius $R$ around an optical source. The halo is assumed to consist of
uniformly distributed HI gas and dust. The optical depth of HI
scattering over a light path $dl$ is $d\tau=\sigma(\nu)n_{\rm
HI}dl$, where $n_{\rm HI}$ is the number density of HI, and
$\sigma(\nu)$ is the cross section of the resonant scattering of
Ly$\alpha$ photons by neutral hydrogen, which is given by
\begin{equation}
\sigma(x)=\sigma_0
\phi(x,a)=\sigma_0\frac{a}{\pi^{3/2}}\int^{\infty}_{-\infty} dy
\frac {e^{-y^2}}{(x-y)^2+a^2}
\end{equation}
where $\phi(x,a)$ is the normalized Voigt profile (Hummer 1965). As
usual, the photon frequency $\nu$ in eq.(1) is described by
the dimensionless frequency $x \equiv (\nu-\nu_0)/\Delta \nu_D$, with
$\nu_0=2.46\times10^{15}$ s$^{-1}$ being the resonant frequency,
$\Delta \nu_D=\nu_0 (v_T/c)= 1.06\times 10^{11}(T/10^4)^{1/2}$ Hz
the Doppler broadening, $v_T=\sqrt{2k_BT/m}$ the thermal
velocity, and $T$ the gas temperature of the halo.
$\sigma_0/\pi^{1/2}$ is the cross section of scattering at the the
resonant frequency $\nu_0$. The parameter $a$ in eq.(1) is the ratio
of the natural to the Doppler broadening. For the Ly$\alpha$ line,
$a=4.7\times 10^{-4}(T/10^4)^{-1/2}$. The optical depth of Ly$\alpha$ photons
with respect to HI resonant scattering is
$\tau_s(x) = n_{\rm HI}R \sigma(x) = \tau_0\phi(x,a)$,
where $\tau_0=n_{\rm HI}\sigma_0R$.

If the absorption and scattering of dust are described by effective
cross-section per hydrogen atom $\sigma_d(x)$, the total optical depth is given by
\begin{equation}
\tau(x) = \tau_0\phi(x,a)+ \tau_{d}(x)
\end{equation}
where the dust optical depth $\tau_{d}(x)=n_{\rm HI}\sigma_{d}(x)R$. This is equal to
assume that dust is uniformly distributed in IGM. The effects of inhomogeneous
density distributions of dust (Neufeld 1991; Haiman \& Spaans 1999) will not be studied
in this paper.

The radiative transfer equation of Ly$\alpha$ photons in a spherical halo with dust
is given by
\begin{eqnarray}
\lefteqn{ {\partial I\over\partial \eta} + \mu \frac{\partial I}
{\partial r}+\frac{(1-\mu^2)}{r}\frac{\partial I}{\partial \mu}
- \gamma \frac{\partial I}{\partial x} = } \nonumber \\
 & &  - \phi(x;a)I + \int \mathcal{R}(x,x';a)I(\eta, r,
x',\mu')dx'd\mu'/2 \\ \nonumber & &-\kappa(x) I + A\kappa(x)\int
\mathcal{R}^d(x,x';\mu,\mu';a)I(\eta, r, x',\mu')dx'd\mu' + S
\end{eqnarray}
where $I(t, r_p, x, \mu)$ is the specific intensity, which is a function of time
$t$, radial coordinate $r_p$, frequency $x$ and  the
direction angle, $\mu=\cos \theta$, with respect to the radial vector ${\bf r}$.

In eq.(3), we use the dimensionless time $\eta$ defined as
$\eta=cn_{\rm HI}\sigma_0 t$
and the dimensionless radial coordinate $r$ defined as $r=n_{\rm
HI}\sigma_0 r_p$. That is, $\eta$ and $r$ are, respectively, in the
units of mean free flight-time and mean free path of photon $\nu_0$ with
respect to the resonant scattering without dust scattering and absorption.
Without resonant scattering, a signal propagates in the radial direction
with the speed of light, the orbit of the signal is then $r= \eta+ {\rm const}$.
With dimensionless variable, the size of the halo $R$ is equal to $\tau_0$.

The re-distribution function $\mathcal{R}(x,x';a)$ gives the
probability of a photon absorbed at the frequency $x'$, and
re-emitted at the frequency $x$. It depends on the details of the
scattering (Henyey \& Greestein 1941; Hummer 1962; Hummer 1969). If we consider
coherent scattering without recoil, the re-distribution function
with the Voigt profile can be written as,
\begin{eqnarray}
\lefteqn{ \mathcal{R}(x,x';a)= } \\ \nonumber
 & \ \ \  & \frac{1}{\pi^{3/2}}\int^{\infty}_{|x-x'|/2}e^{-u^2}
\left [
\tan^{-1}\left(\frac{x_{\min}+u}{a}\right)-\tan^{-1}\left(\frac{x_{\max}-u}{a}\right
)\right ]du
\end{eqnarray}
where $x_{\min}=\min(x, x')$ and $x_{\max}=\max(x,x')$. In the case
of $a=0$, i.e. considering only the Doppler broadening, the
re-distribution function is
\begin{equation}
\mathcal{R}(x,x')=\frac{1}{2}{\rm erfc}[{\rm max}(|x|,|x'|)].
\end{equation}
The re-distribution function of equation (5) is normalized as
$\int_{-\infty}^{\infty} \mathcal{R}(x,x')dx'=\phi(x,0)
=\pi^{-1/2}e^{-x^2}$. With this normalization, the total number of
photons is conserved in the evolution described by equation (3).
That is, the destruction processes of Ly$\alpha$ photons, such as
the two-photon process (Spitzer \& Greenstein 1951; Osterbrock
1962), are ignored in equation (3). The recoil of atoms is also not considered
in equation (4) or (5). The effect of recoil actually is under control
(Roy et al. 2009c, 2010). We will address it in next section.

The absorption and scattering of dust are described by the term $\kappa(x)I$ of eq.(3),
where $\kappa(x)=\sigma_{d}/\sigma_0$, which is of the order of
$ 10^{-8}(T/10^4)^{1/2}$ (Draine \& Lee 1984; Draine 2003).
The term with $A$ of eq.(3) describes albedo, i.e.
$A\equiv \sigma_s/\sigma_d$, where $\sigma_s$ is the cross section of dust
scattering. Generally, $A$ lies approximately between 0.3 and 0.4 (Pei 1992;
Weingartner \& Draine 2001).

Since dust generally is much heavier than a single atoms, the recoil of dust particles can
be neglected when colliding with a photon. Under this ``heavy dust'' approximation,
photons do not change their frequency during the collision with dust. The redistribution
function of dust
$\mathcal{R}^d$ is independent of $x$ and $x'$, and is simply given by a phase function
as
\begin{equation}
\mathcal{R}^d(\mu,\mu')  =  \frac{1}{4\pi}\int^{2\pi}_{0}
   d\phi' \frac{1-g^2}{(1+g^2-2g\bar{\mu})^{3/2}}
  =\sum_{l=0}^{\infty}\frac{(2 l+1)}{2} g^l P_l(\mu)P_l(\mu'),
 \end{equation}
where $\bar{\mu}=\mu\mu'+\sqrt{(1-\mu^2)(1-\mu'^2)}\rm{cos}\phi'$ and $P_l$ is the
Legendre function. The factor
$g$ in eq.(6) is the asymmetry parameter. For isotropic scattering,
$g=0$. The cases of $g=+1$ and -1 correspond to complete forward and backward
scattering, respectively. Generally, the factor $g$ is a function
of the wavelength. For the Ly$\alpha$ photon, we will take $g=0.73$
for realistic dust scattering (Li \& Draine 2001). The integral of eq.(6) is
performed in Appendix A.

In eq. (3), the term with the parameter $\gamma$ is due to the
expansion of the universe. If $n_{\rm H}$ is equal to the mean of
the number density of cosmic hydrogen, we have $\gamma=\tau_{GP}^{-1}$, and
$\tau_{GP}$ is the Gunn-Peterson optical depth. Since the Gunn-Peterson optical
depth is of the order of $10^{6}$ at high redshift (e.g. Roy et al. 2009c),
the parameter $\gamma$ is of the order of $10^{-5}-10^{-6}$. Therefore, if the
optical depth of halos is equal to or less than 10$^6$, the term with $\gamma$
of eq.(3) can be ignored.

In eq.(3) we neglect the effect of collision transition from $H(2p)$
state to $H(2s)$ state, which can significantly affect on the escape
of Ly$\alpha$ photons when HI column density is higher than
$10^{21}$ cm$^{-2}$ and dust absorption is very small (Neufeld,
1990). This generally is out of the parameter range used below. We
are also not considering the effects of bulk motion of the medium of
halos (e.g. Spaans \& Silk 2006, Xu \& Wu, 2010).

\subsection{Eddington approximation}

Eq.(6) indicates that the transfer equation (3) can be solved with
the Legendre expansion $I(\eta,r,x,\mu)= \sum_{l}I_l(\eta,r,x)P_l(\mu)$. If we take
only the first two terms, $l=0$ and 1, it is the Eddington approximation as
\begin{equation}
I(\eta,r,x,\mu)\simeq J(\eta, r, x) + 3\mu F(\eta, r,x)
\end{equation}
where
\begin{equation}
J(\eta,r,x)=\frac{1}{2}\int_{-1}^{+1}I(\eta,r,x,\mu)d\mu,  \hspace{5mm}
F(\eta,r,x)=\frac{1}{2}\int_{-1}^{+1}\mu I(\eta,r,x,\mu)d\mu.
\end{equation}
They are, respectively, the angularly averaged specific intensity and flux. Defining
$j=r^2J$ and $f=r^2F$, Eq.(3) yields the equations of $j$ and $f$ as
\begin{eqnarray}
{\partial j\over\partial \eta} + \frac{\partial f} {\partial r} & =
& - (1-A)\kappa j- \phi(x;a)j + \int \mathcal{R}(x,x';a)j dx'+ \gamma
\frac{\partial
j}{\partial x}+ r^2S,\\
 \frac{\partial f}{\partial \eta} + \frac{1}{3}
\frac{\partial j} {\partial r} - \frac{2}{3}\frac{j}{r} & = & -(1-Ag)\kappa f +\gamma
\frac{\partial
f}{\partial x}-
\phi(x;a)f.
\end{eqnarray}
The mean intensity $j(\eta,r,x)$ describes the $x$ photons
trapped in the position $r$ at time $\eta$ by the resonant scattering, while the flux
$f(\eta,r,x)$ describes the photons in transit.

The source term $S$ in the equations (3) and (9) can be described by a
boundary condition of $j$ and $f$ at $r=r_0$. We can take $r_0=0$. Thus, the
boundary condition is
\begin{equation}
j(\eta, 0, x)=0, \hspace{1cm} f(\eta, 0, x)=S_0\phi_s(x),
\end{equation}
where $S_0$, and $\phi_s(x)$ are, respectively, the intensity and
normalized  frequency profile of the sources. Since equation
(3) is linear, the solutions of $j(x)$ and $f(x)$ for given $S_0=S$ are equal to
$S j_1(x)$ and $S f_1(x)$, where $j_1(x)$ and $f_1(x)$ are the solutions
of $S_0=1$. On the
other hand, the equation (3) is not linear with respect to the function $\phi_s(x)$.
The solution $f(x)$ for a given $\phi_s(x)$ is not equal to
$\phi_s(x)f_1(x)$, where $f_1(x)$ is the solution of $\phi_s(x)=1$.

In the range outside the halo, $r>R$, no photons propagate in the direction
$\mu<0$. The boundary condition at $r=R$ given by
$\int_{0}^{-1}\mu J(\eta, R, x, \mu)d\mu =0$ is then (Unno 1955)
\begin{equation}
j(\eta, R,x)=2f(\eta, R,x).
\end{equation}
There is no photon in the field before $t=0$. Therefore, the initial
condition is
\begin{equation}
j(0,r,x)=f(0,r,x)=0.
\end{equation}
We will solve equations (9) and (10) with boundary and initial conditions eqs.(11) - (13)
by using the WENO solver (Roy et al. 2009a, b, c, 2010). Some details of this
method is given in Appendix B.

\subsection{Dust models}

We consider three models of the dust as follows:
\begin{description}
\item I.  pure scattering, $A=1, \ g=0.73$: dust causes only anisotropic scattering, but no absorption;
\item II. scattering and absorption. $A=0.32, \ g=0.73$: dust causes both absorption and
 anisotropic scattering.
\item III. pure absorption. $A=0$: dust causes only absorption, but no scattering;
\end{description}
Models I and III do not occur in reality. They are, however, helpful to reveal the effects
of pure scattering and absorption on the radiative transfer.

Since $\kappa(x)$ is on the order of $10^{-8}$, its effect will be significant only for
halos with halos with optical depth $\tau_0 \geq 10^6$, and ignorable for
$\tau_0 \leq 10^5$. To illustrate the dust effect, we use halos of $R=\tau_0 \leq 10^4$, and take larger $\kappa$ to be
$\simeq 10^{-4}-10^{-2}$. We also assume that $\kappa$ is frequency-independent.
We consider below only the case of grey dust, i.e. $\kappa$ is independent of frequency $x$.
This certainly is not realistic dust. Yet, the frequency range given in solution below mostly are
in the range $|x|< 4$. Therefore, the approximation of grey dust would be proper if cross section
of dust is not significantly frequency dependent in the range  $|x|< 4$.

\subsection{Numerical example: Wouthuysen-Field thermalization}

As the first example of numerical solutions, we show the
Wouthuysen-Field (W-F) effect, which requires that the distribution
of Ly$\alpha$ photons in the frequency space should be thermalized
near the resonant frequency $\nu_0$. The W-F effect illustrates the
difference between the analytical solutions of the Fokker-Planck
approximation and that of eqs. (9) and (10). The former can not show
the local thermalization (Neufeld 1990), while the latter can (Roy
et al. 2009b). All problems related to the W-F local thermal
equilibrium should be studied with the integro-differential equation
(3).

\begin{figure}[htb]
\begin{center}
\includegraphics[scale=0.20]{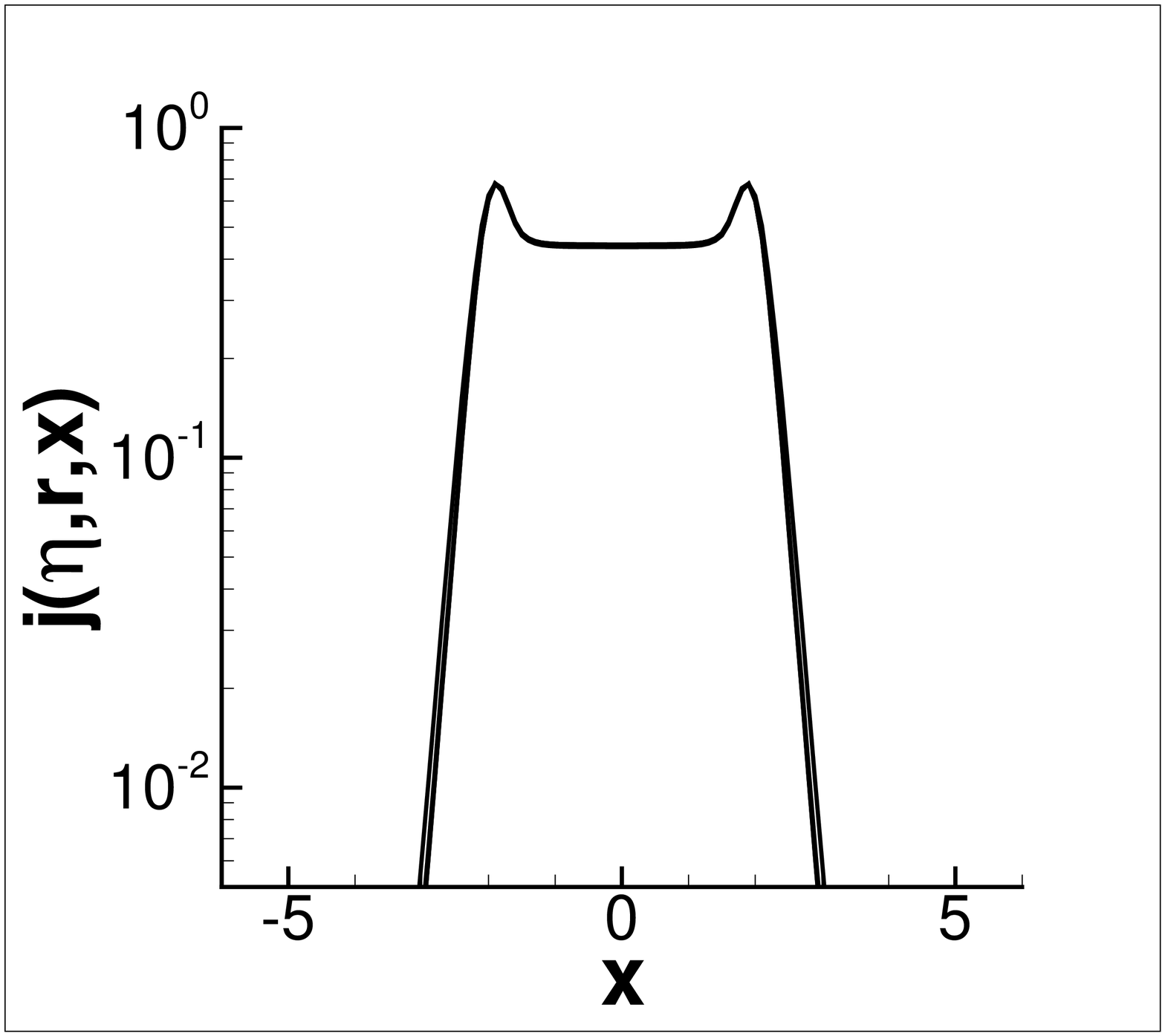}
\includegraphics[scale=0.20]{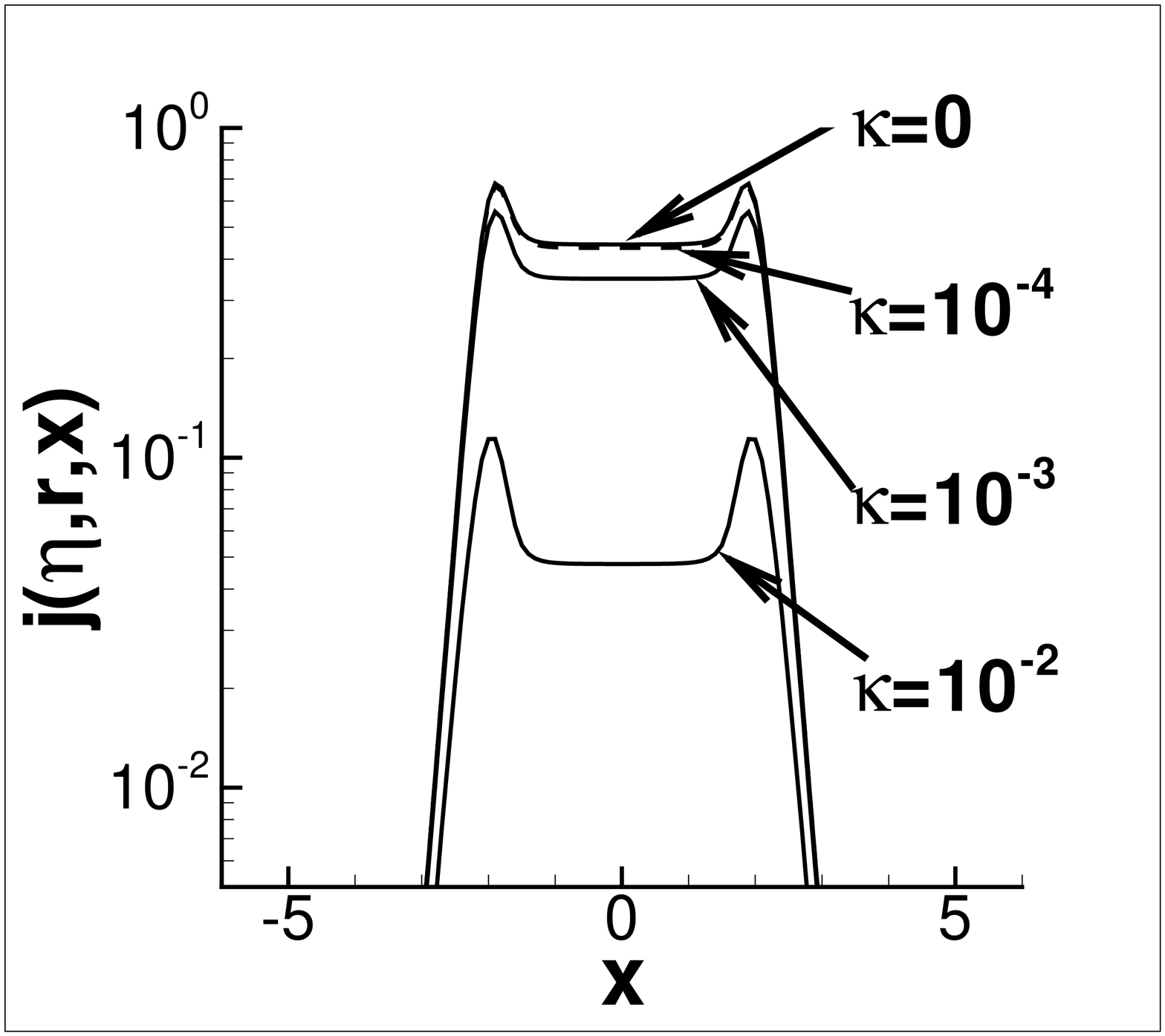}
\includegraphics[scale=0.20]{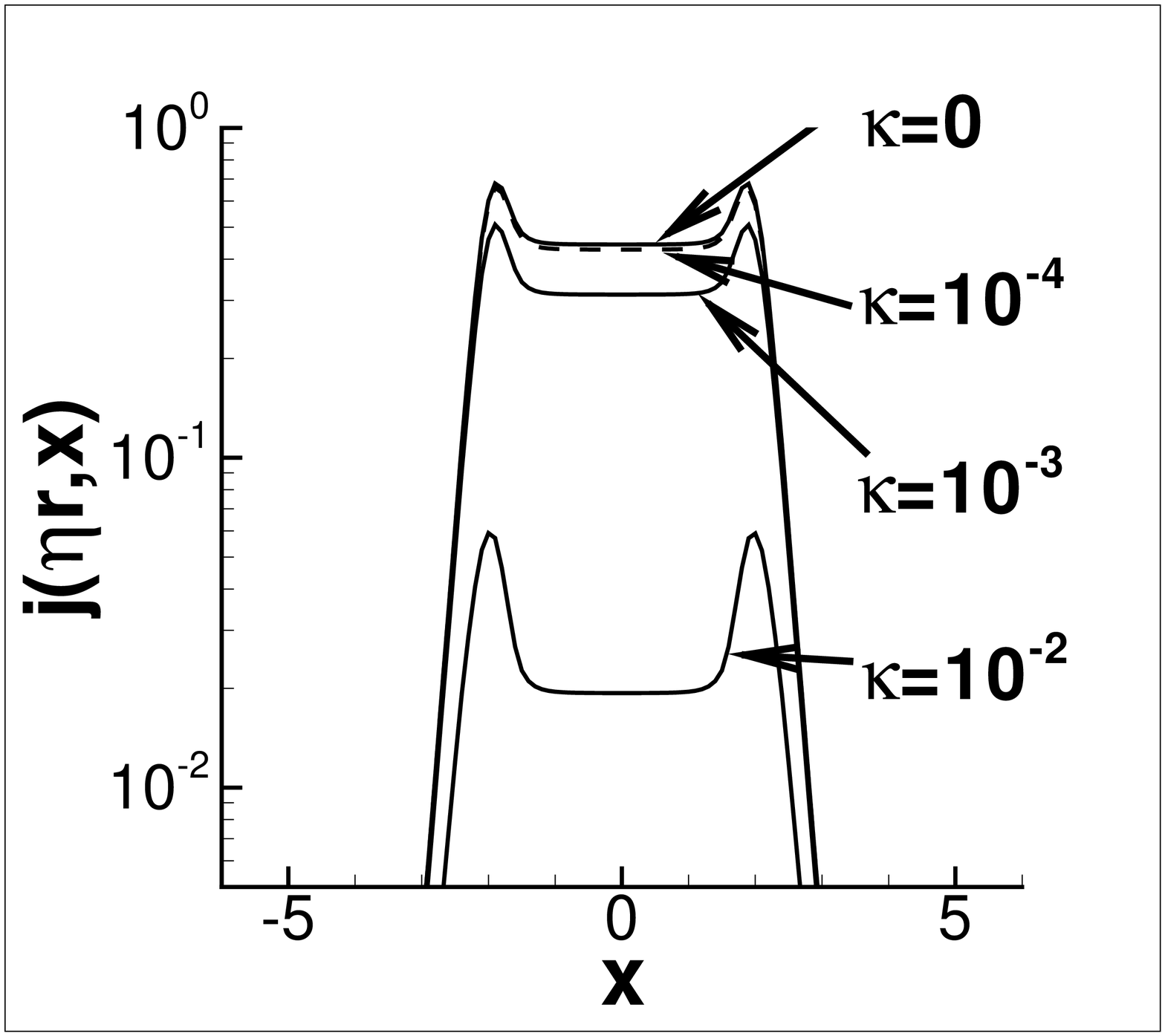}
\end{center}
\caption{The mean intensity $j(\eta,r,x)$ at $\eta=500$ and $r=100$
for dust models I (left panel), II (middle panel) and III (right
panel). The source is $S_0=1$ and
$\phi_s(x)=(1/\sqrt{\pi})e^{-x^2}$. The parameter $a=10^{-3}$. In
each panel, $\kappa$ is taken to be 0, 10$^{-4}$, 10$^{-3}$ and
10$^{-2}$.}
\end{figure}

Figure 1 presents a solution of mean intensity $j(\eta,r,x)$
at time radial $\eta=500$ coordinate $r=10^2$ for halo with size $R\gg r=10^2$.
The three panels correspond to dust models I (left panel), II (middle panel)
and III (right panel). The source is
taken to have a Gaussian profile $\phi_s(x)=(1/\sqrt{\pi})e^{-x^2}$
and unit intensity $S_0=1$. The solutions of Figure 1 actually are independent of $R$,
if $R\gg 10^2$. The intensity of $j$ is
decreasing from left to right in Figure 1, because the absorption is
increasing with the models from I to III.

A remarkable feature shown in Figure 1 is that all $j(\eta,r,x)$
have a flat plateau in the range $|x|\leq 2$. This gives the frequency range of
the W-F local thermalization (Roy et al, 2009b, c). The range of the flat plateau
$|x|\leq 2$ is almost dust-independent, either for model I or for models
II and III. This is expected, as neither the absorption nor scattering
given by the $\kappa$ term of eq.(3) changes the frequency
distribution of photons. The redistribution function (6) also does not
change the frequency distribution of photons. This point can also be seen
from eqs.(9) and (10), in which the $\kappa$ terms are frequency-independent.
The evolution of the frequency distribution of photons is due only to the
resonant scattering.

Since thermalization will erase all frequency features within the range
$|x|\leq 2$, the double-peaked structure does not retain information of
the photon frequency distribution within $|x|<2$ at the source. That is,
the results in Figure 1 will hold for any source $S_0\phi_s(x)$ with arbitrary
$\phi_s(x)$ which is non-zero within  $|x|<2$ (Roy et al. 2009b, c).
This property can also be used as a test of the simulation code.
It requires that simulation results of flat plateau should be hold, regardless
of the source to be monochromatic or with finite width around $\nu_0$.

\section{Dust effects on photon escape}

\subsection{Model I: scattering of dust}

To study the effects of dust scattering on the Ly$\alpha$ photon
escape, we show in Figure 2 the flux $f(\eta,r,x)$ of Ly$\alpha$
photons emergent from halos at the boundary $r=R=10^2$ for Model I. The
three panels of Figure 2 correspond to
$\kappa=10^{-4}$, 10$^{-3}$, and 10$^{-2}$ from left to right,
respectively. The source starts to emit photons at $\eta=0$
with a stable luminosity $S_0=1$, and with a Gaussian profile
$\phi_s(x)=(1/\sqrt{\pi})e^{-x^2}$.

\begin{figure}[htb]
\begin{center}
\includegraphics[scale=0.20]{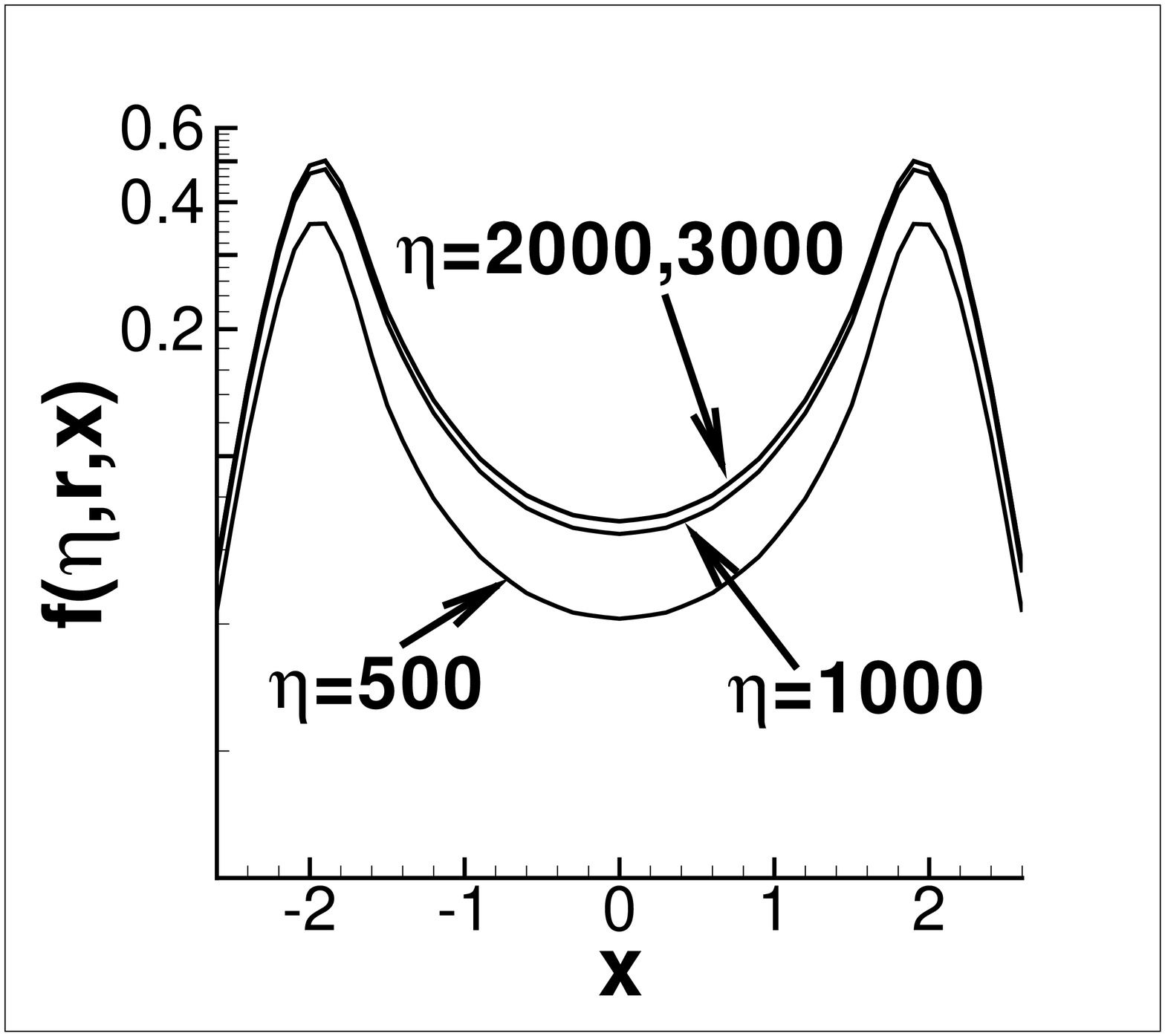}
\includegraphics[scale=0.20]{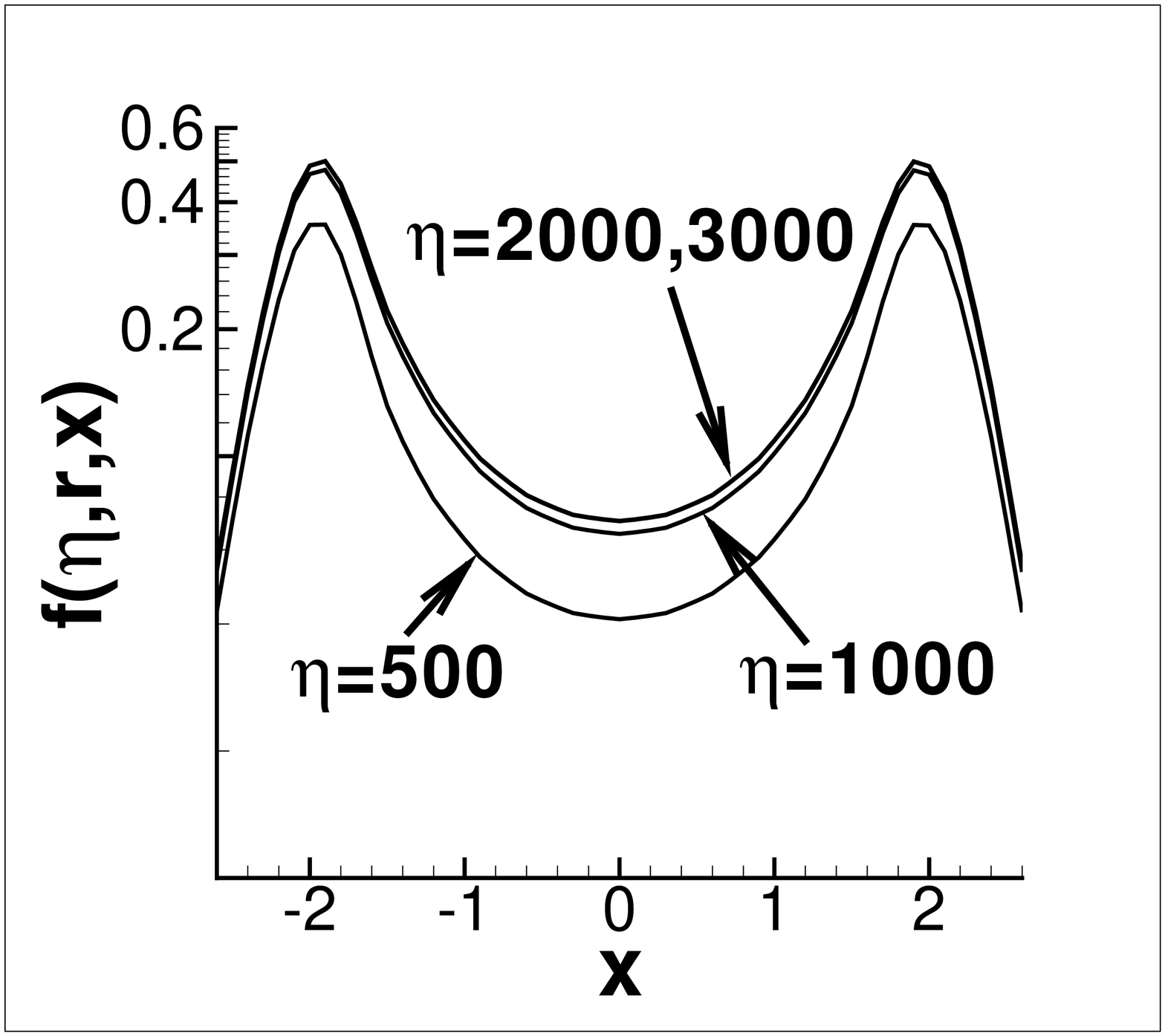}
\includegraphics[scale=0.20]{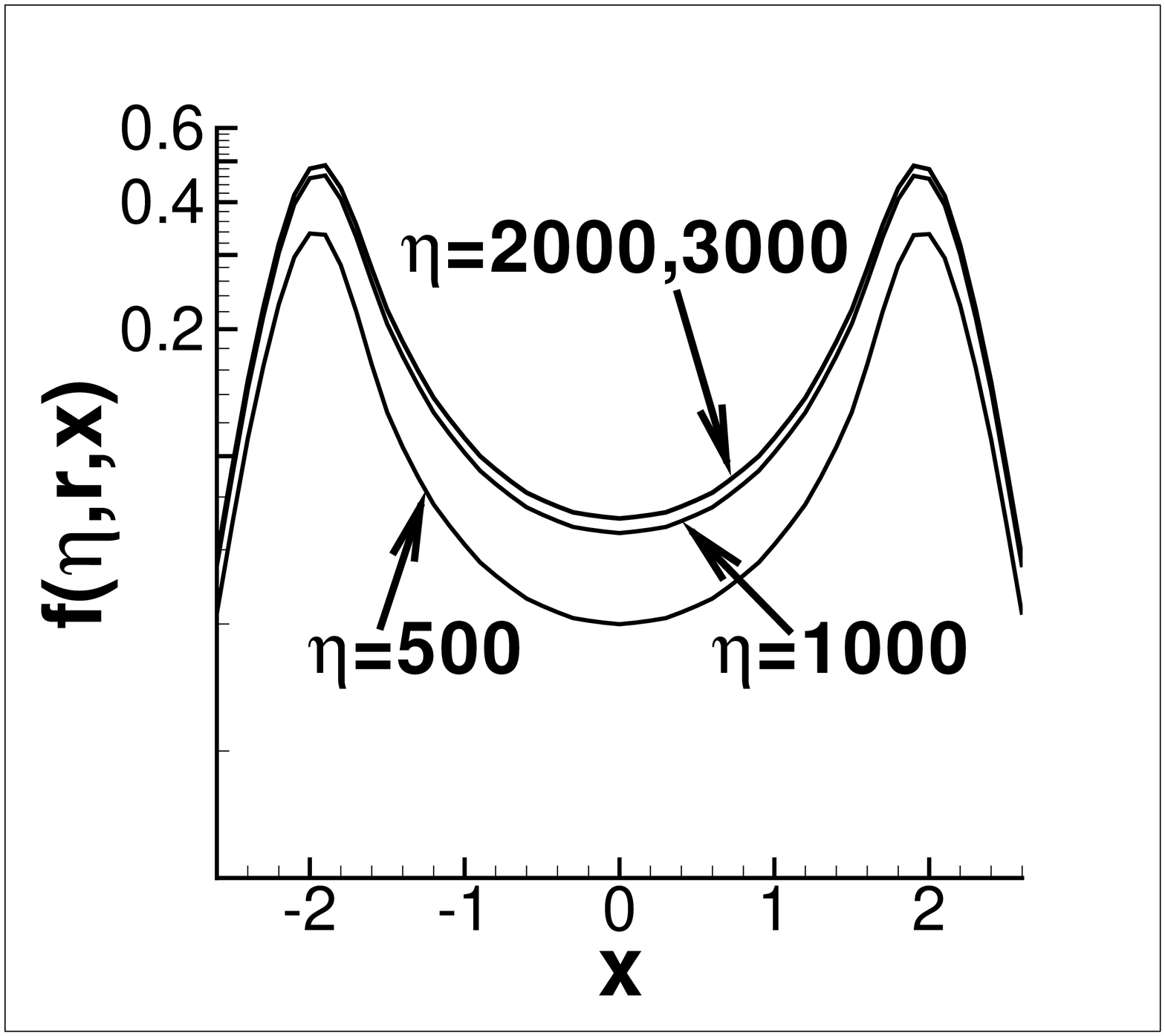}
\end{center}
\caption{Flux $f(\eta,r,x)$ of Ly$\alpha$ photons emergent from
halos at the boundary $R=10^2$, and for the dust model I $A=1, \ g=0.73$.
The parameter $\kappa$ is taken to be 10$^{-4}$ (left),
10$^{-3}$(middle) and 10$^{-2}$ (right). The source is $S_0=1$ and
$\phi_s(x)=(1/\sqrt{\pi})e^{-x^2}$.  The parameter $a=10^{-3}$.}
\end{figure}

Figure 2 clearly shows that the time-evolution of $f(\eta,r,x)$ is
$\kappa$-independent. Although the cross section of dust scattering
increases about 100 times from $\kappa=10^{-4}$ to $\kappa=10^{-2}$,
the curves of the left and right panels in Figure 2 actually are almost
identical.

According to the scenario of ``single longest excursion'', photon escape
is not a process of Brownian random walk in the spatial space, but a transfer
in the frequency space (Osterbrock 1962; Avery \& House 1968; Adams, 1972, 1975;
Harrington 1973; Bonilha et al. 1979). Photon will escape, once its frequency
is transferred from $|x|< 2$ to $|x|>2$, on which the medium is transparent.
On the other hand, dust scattering given by the redistribution function eq.(6)
does not change photon frequency. Dust scattering has no effect on the transfer
in the frequency space.

Moreover, photons with frequency $|x|<2$ are quickly thermalized after a few
hundred resonant scattering. In the local thermal equilibrium state, the
angular distribution of photons is isotropic. Thus, even if the dust scattering
is anisotropic $g\neq 0$ with respect to the direction of the incident particle,
the angular distribution will keep isotropic undergoing a $g\neq 0$ scattering. Hence,
dust scattering also has no effect on the angular distribution.

\subsection{Model III: absorption of dust}

Similar to Figure 2, we present in Figure 3 the flux of Model III,
i.e. dust causes only absorption without scattering. All other
parameters of Figure 3 are the same as in Figure 2. In the left
panel of Figure 3, the curves at the time $\eta=2000$ and $3000$ are
the same. It means the flux $f(\eta,R,x)$ at the boundary $R$ is
already stable, or saturated at the time $\eta \geq 2000$. The small
difference between the curves of $\eta=1000$ and $\eta \geq 2000$ of
the left panel indicates that the flux is still not yet completely
saturated at the time $\eta=1000$. However, comparing the middle and
right panels of Figure 3, we see that for $\kappa=10^{-3}$, the flux
has already saturated at $\eta = 1600$, while it has saturated at
$\eta=800$ for $\kappa=10^{-2}$. That is, the stronger the dust
absorption, the shorter the saturation time scale. The time scales
of escape or saturation do not increase by dust absorption, and even
decrease with respect to the medium without dust. Stronger
absorption leads to shorter time scale of saturation.

\begin{figure}[htb]
\begin{center}
\includegraphics[scale=0.20]{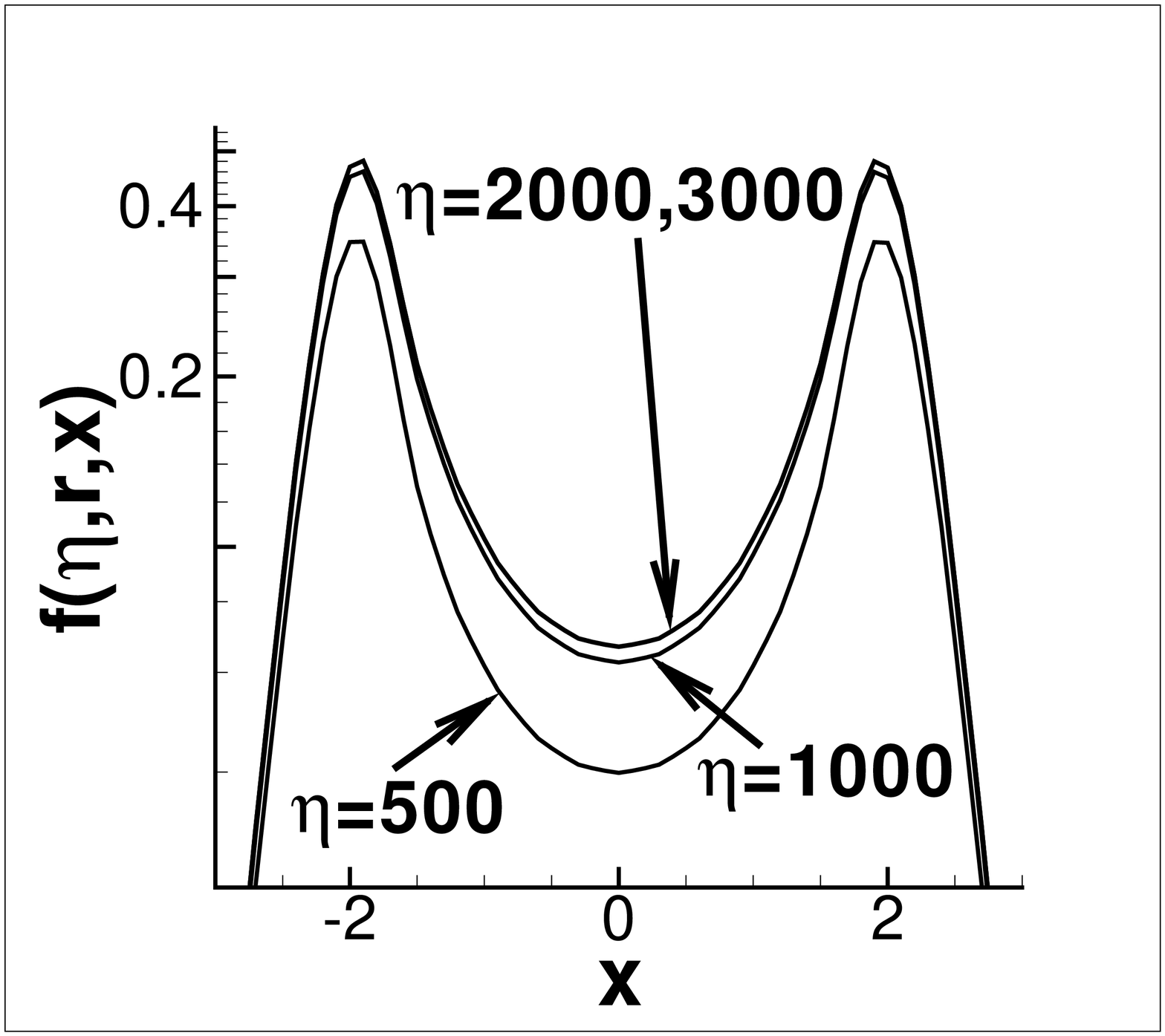}
\includegraphics[scale=0.20]{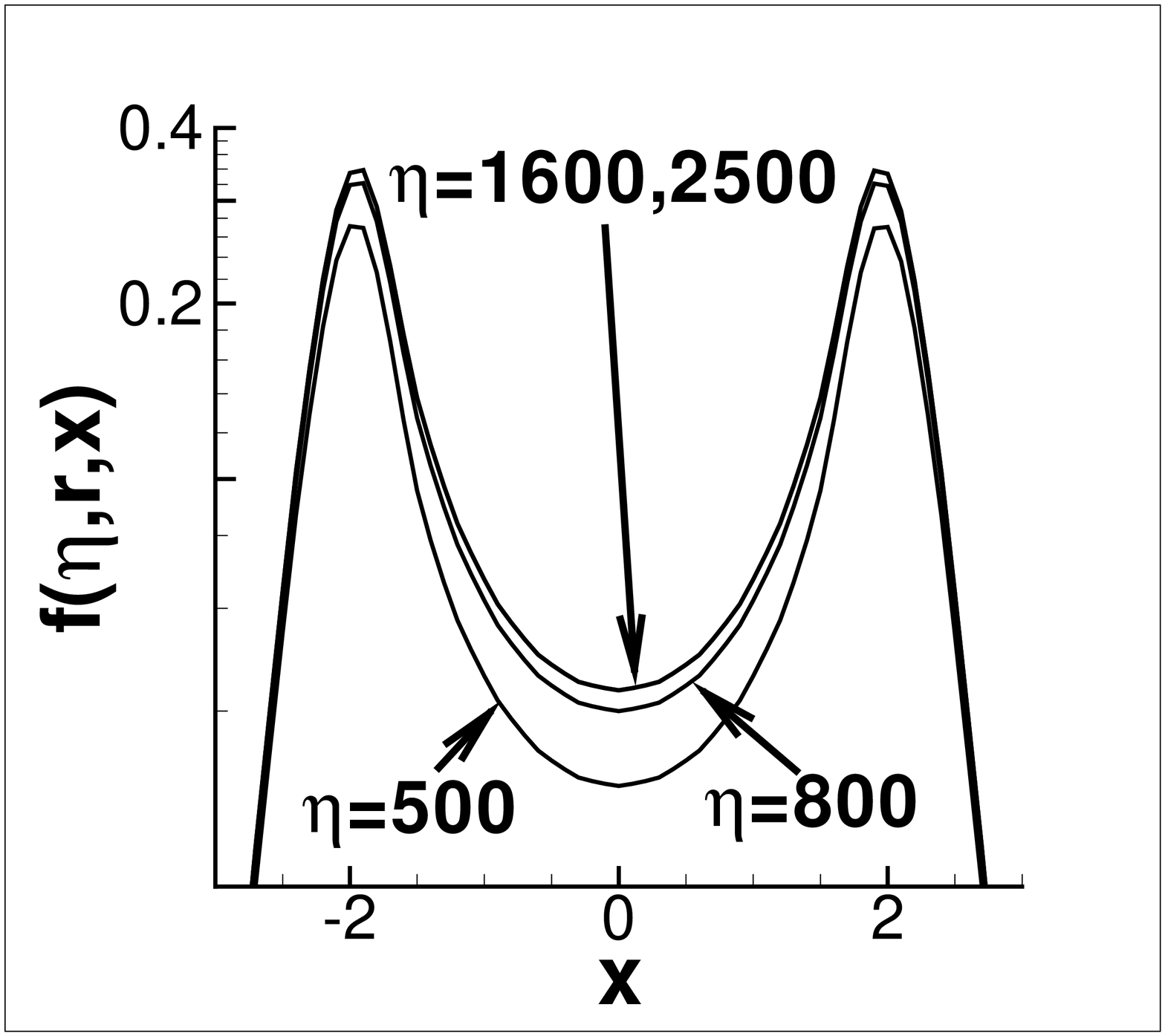}
\includegraphics[scale=0.20]{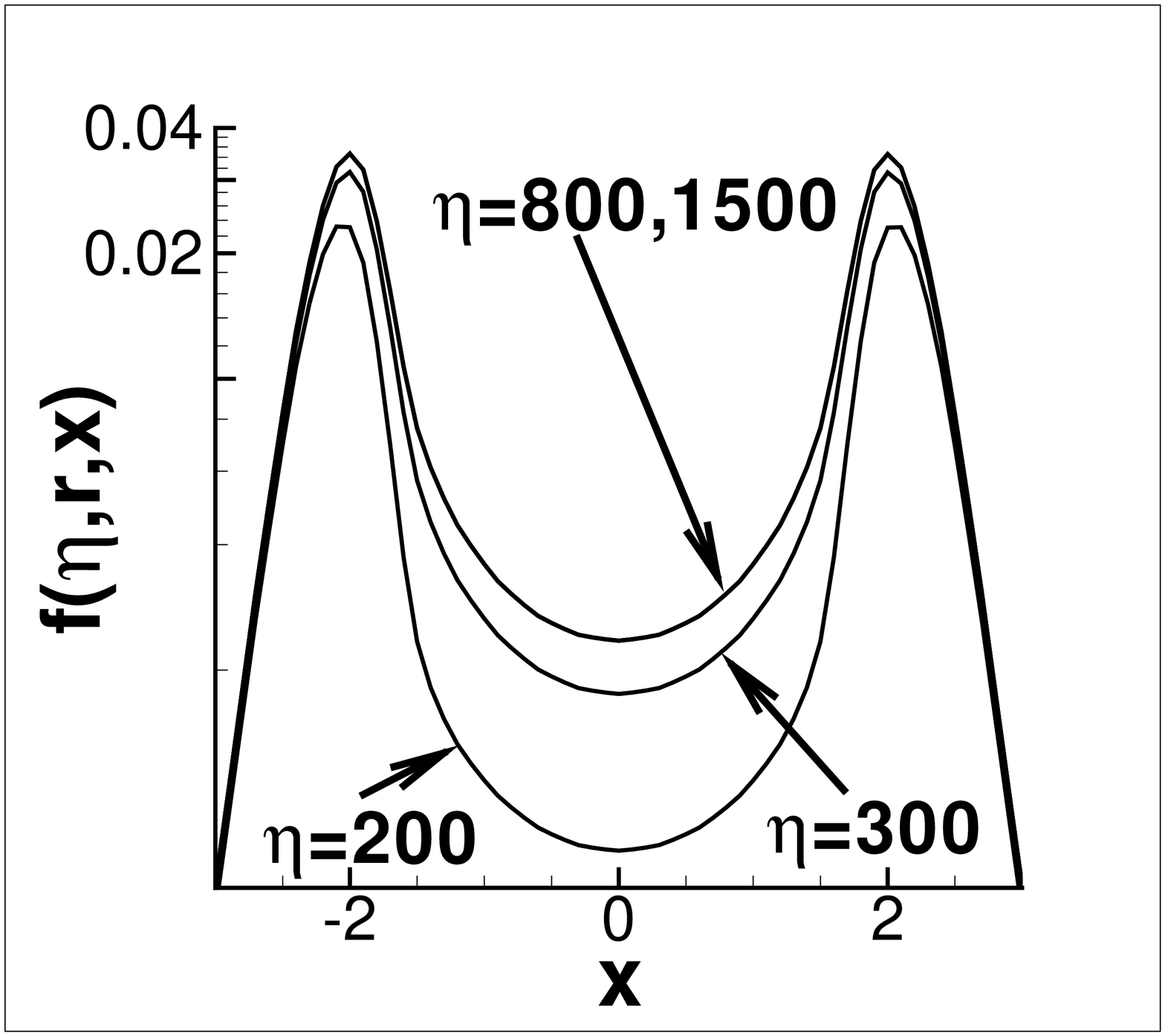}
\end{center}
\caption{Flux $f(\eta,r,x)$ of Ly$\alpha$ photons emergent from
halos at the boundary $r=R=10^2$. The parameters of the dust are $A=0$ and $\kappa=$
10$^{-4}$ (left), 10$^{-3}$ (middle) and 10$^{-2}$ (right).
Other parameters are the same as in Figure 2.}
\end{figure}

Obviously, dust absorption does not help in producing photons for
the ``single longest excursion''. Therefore, dust absorption can not
make the time scale of producing photons for
``single longest excursion'' to be smaller. However, dust
absorptions are effective in reducing the number of photons trapped
in the state of local thermalized equilibrium $|x|<2$ (see also \S 4.2).
This leads to the fact that the higher the value of $\kappa$, shorter the time scale of
saturation.

\subsection{Effective absorption optical depth}

Since Ly$\alpha$ photons underwent a large number of resonant scattering
before escaping from halo with optical depth $\tau_0\gg 1$, it is
generally believed that a small absorption of dust will lead to a significant
decrease of the flux. However, it is still unclear what the exact relationship
between the
dust absorption and the resonant scattering is. This problem should be measured
by the effective optical depth of dust absorption of Ly$\alpha$ photons in
$R=\tau_0\gg 1$ halos.

To calculate the  effective optical depth, we first give the total flux of
Ly$\alpha$ photons emergent from halo of radius $R$, which is defined as
$F(\eta)=\int f(\eta, R, x) dx$. Figure 4 plots
$F(\eta)$ as a function of time $\eta$ for halo with sizes $R=\tau_0=10^2$ and $10^4$. The
curves typically are the time-evolution of growing and then
saturating. The three panels correspond to the dust models I, II and
III from left to right. The upper panels are of $R=10^2$, and lower panels for
$R=10^4$. In each panel of $R=10^2$, we have three curves corresponding to $\kappa=$
10$^{-4}$, 10$^{-3}$ and 10$^{-2}$, respectively. In cases of $R=10^4$, we take $\kappa=$
10$^{-4}$ and 10$^{-3}$.
\begin{figure}[htb]
\begin{center}
\includegraphics[scale=0.20]{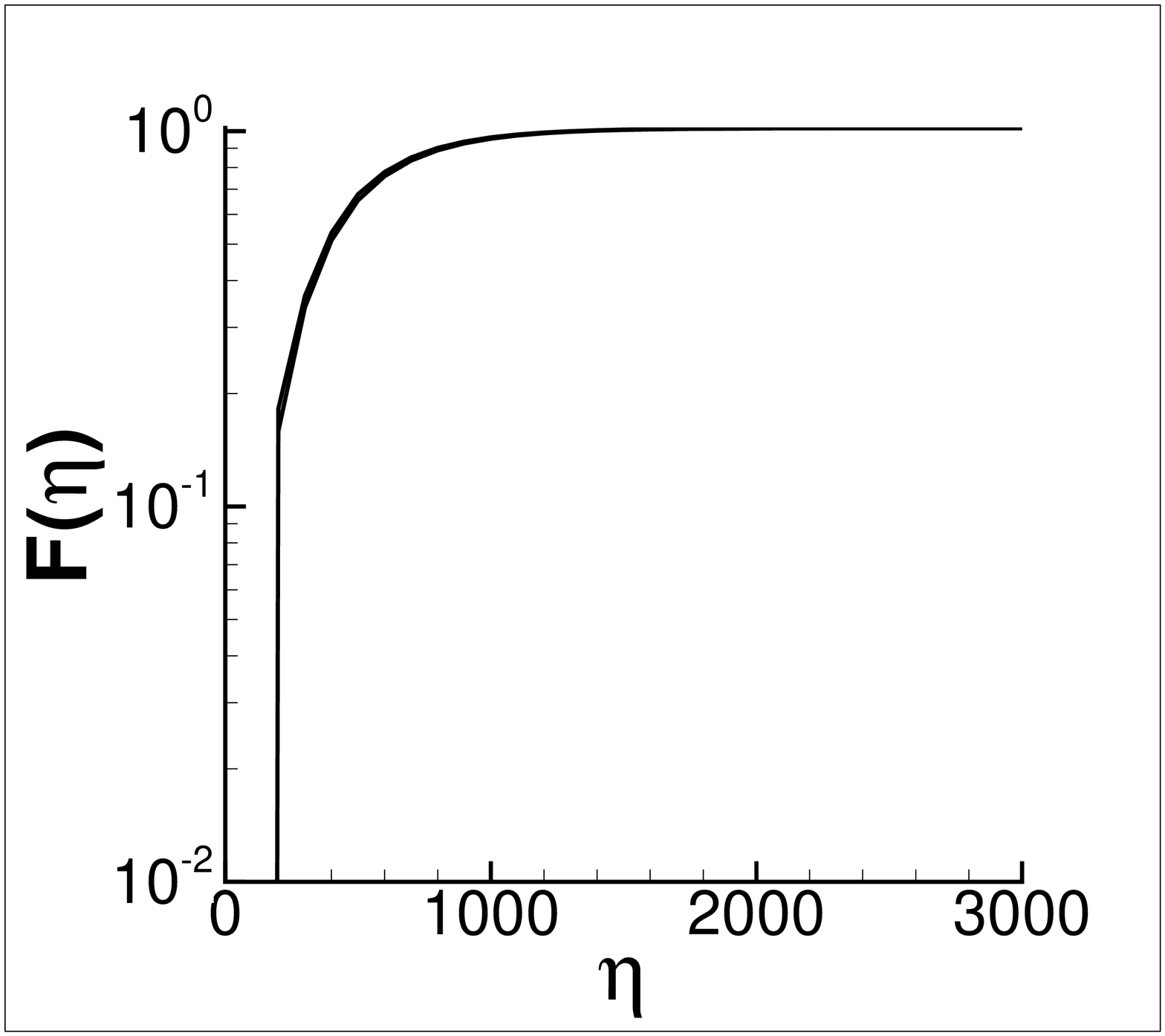}
\includegraphics[scale=0.20]{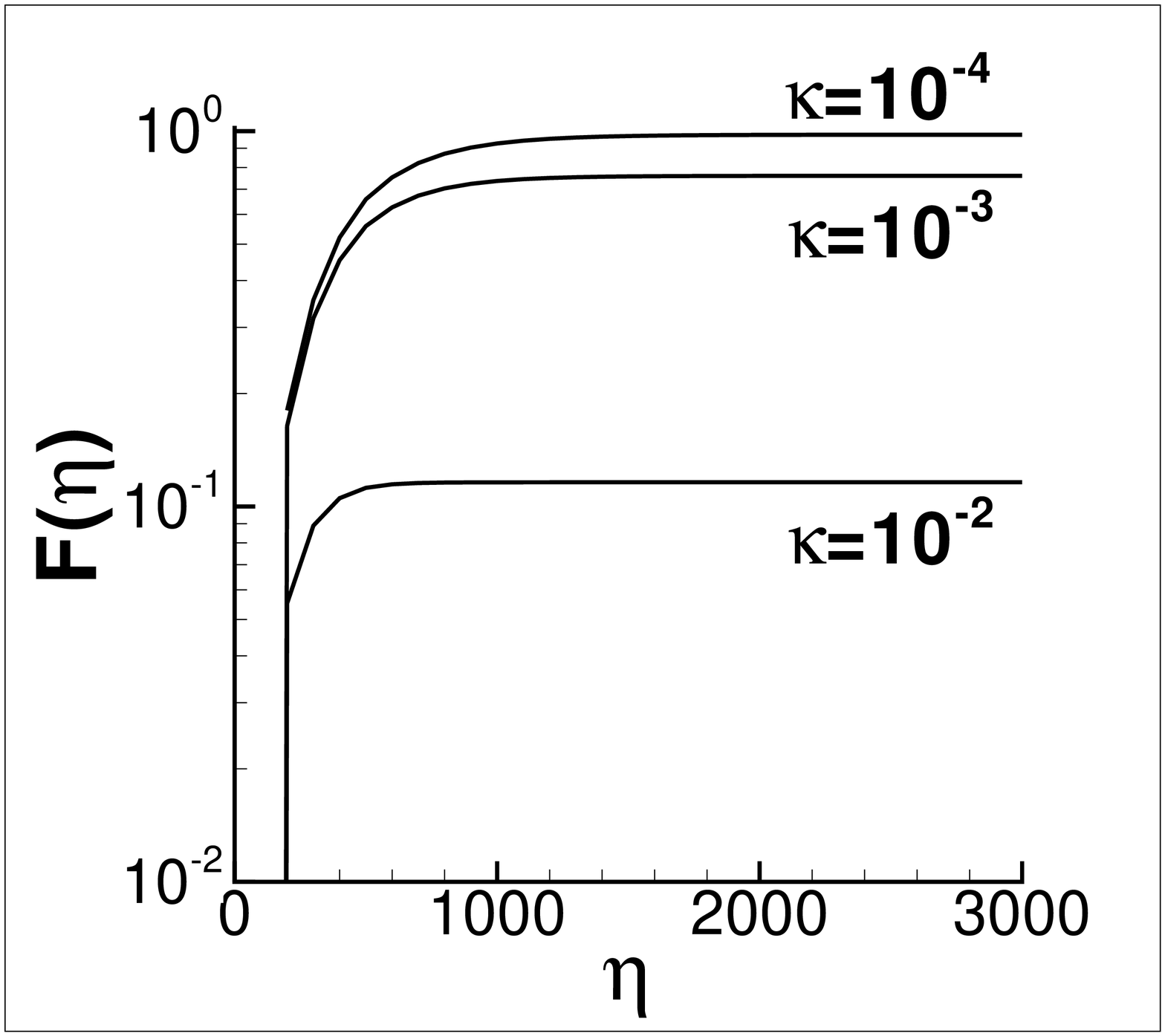}
\includegraphics[scale=0.20]{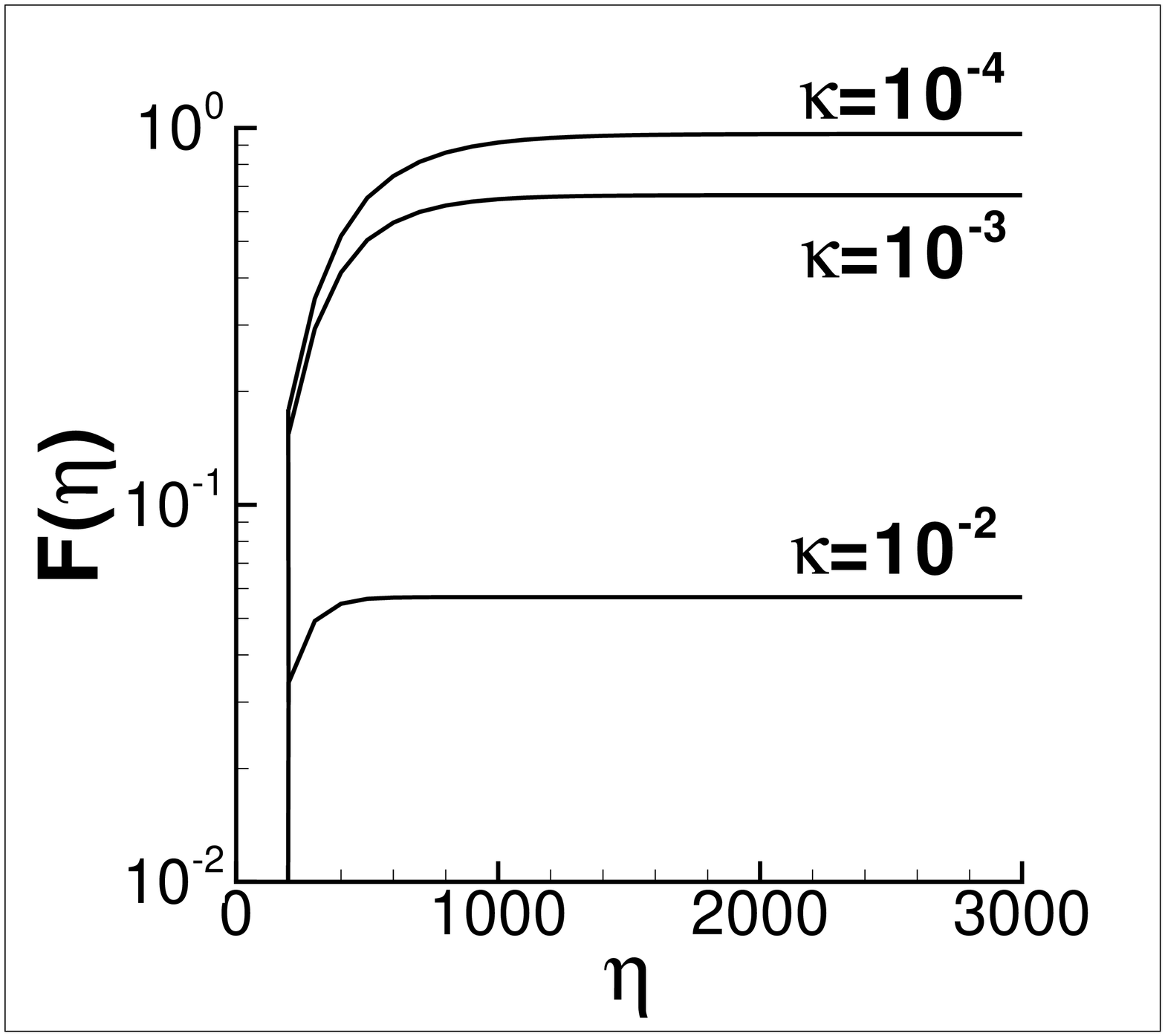}
\end{center}
\begin{center}
\includegraphics[scale=0.20]{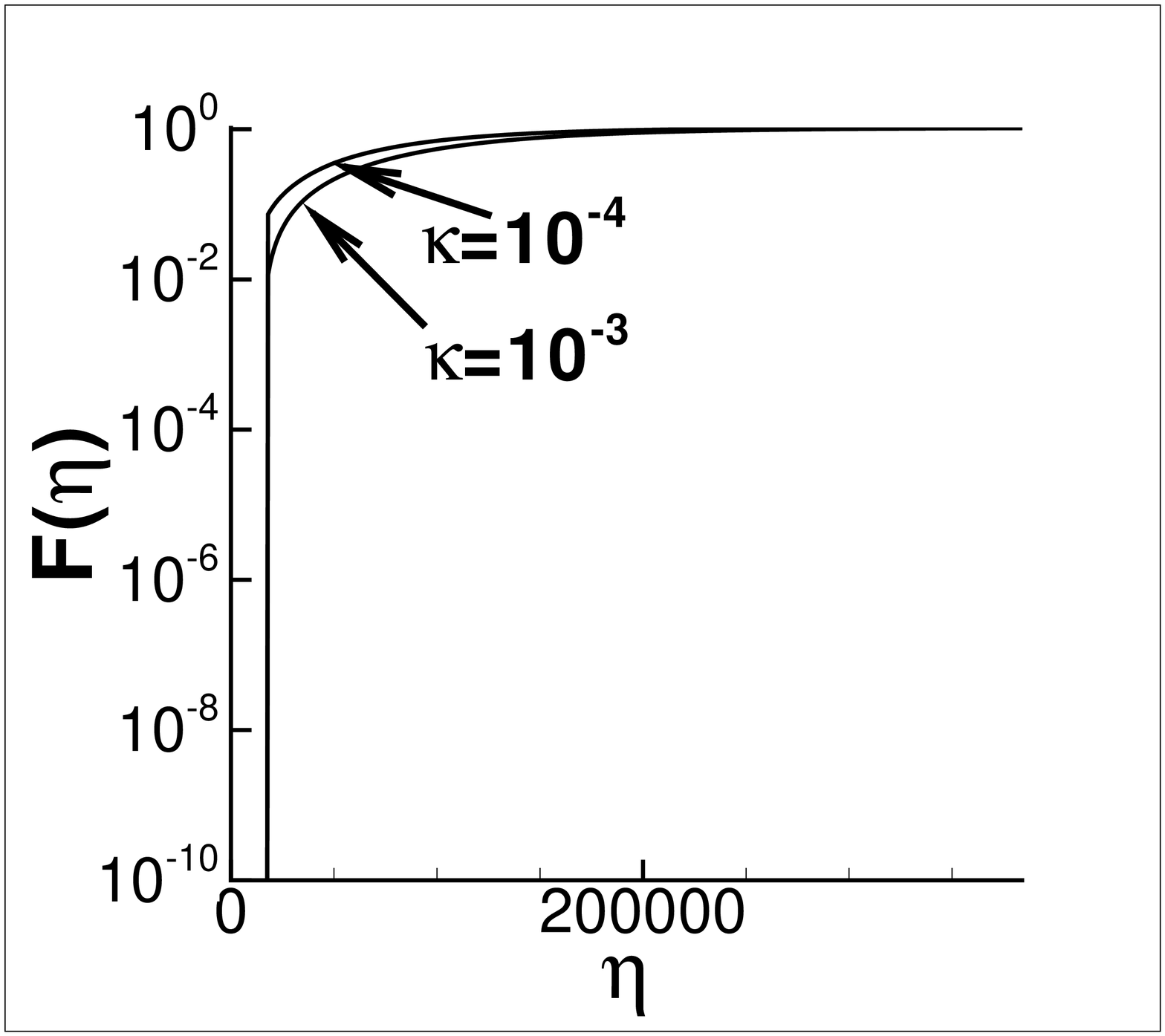}
\includegraphics[scale=0.20]{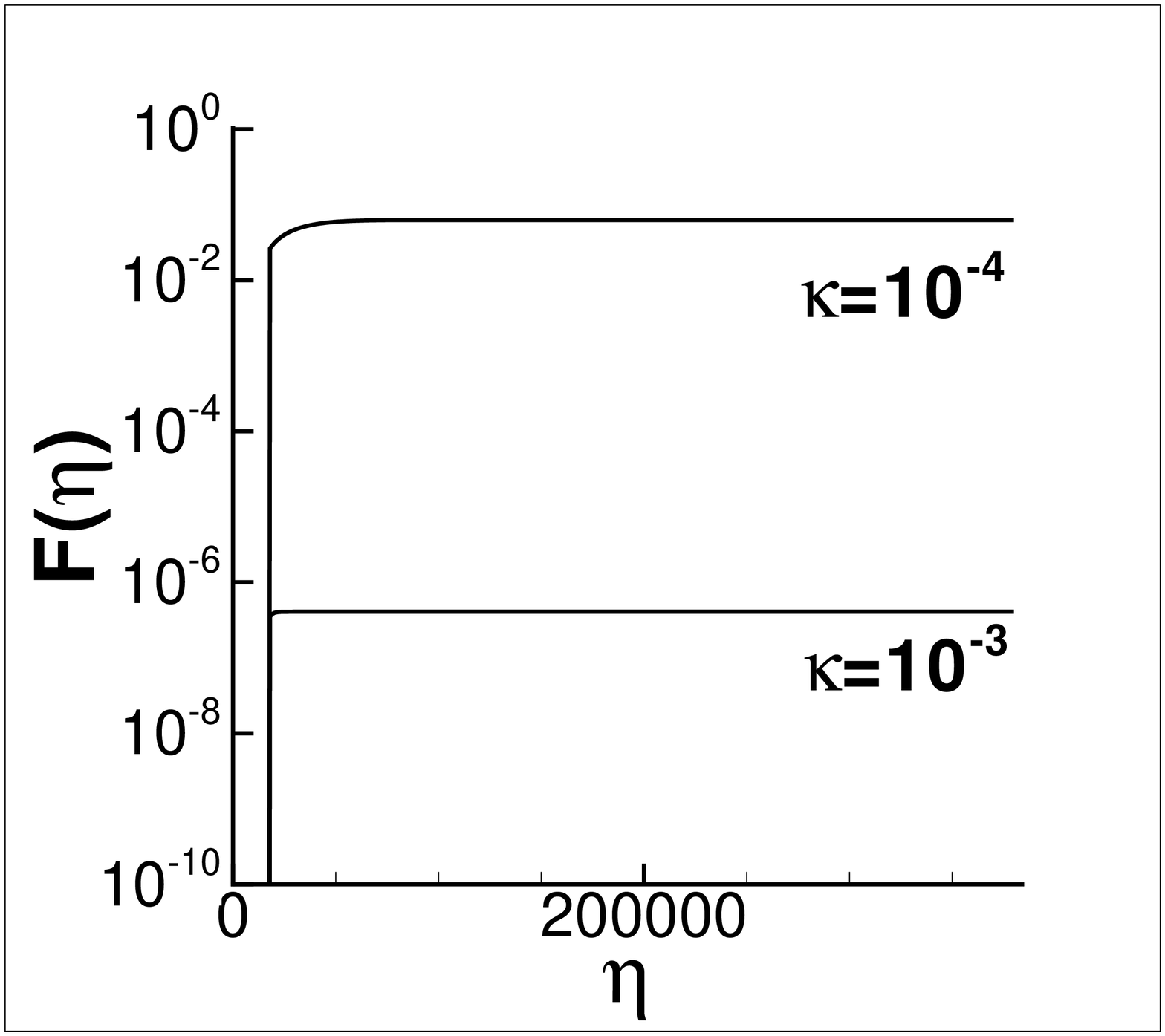}
\includegraphics[scale=0.20]{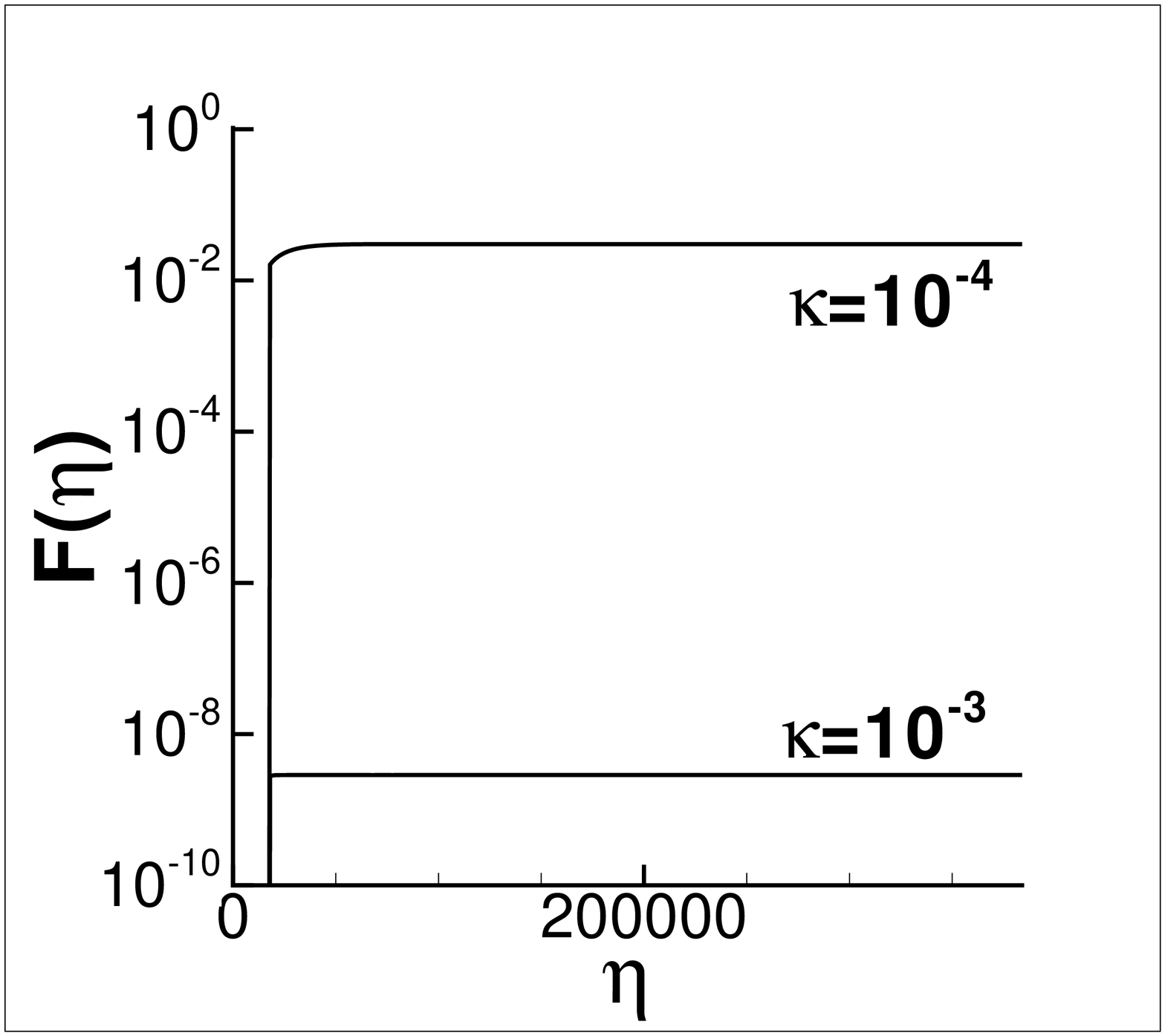}
\end{center}
\caption{The time evolution of the total flux $F(\eta)$ at the boundary of halos with
$R=\tau_0=10^2$ (upper panels), and $R=\tau_0=10^4$ (lower Panels). The source of $S_0=1$ and
$\phi_s(x)=(1/\sqrt{\pi})e^{-x^2}$ starts to emit photons at
time $\eta=0$. The parameters of dust are $(A=1, g=0.73)$ (left);
$(A=0.32, \ g=0.73)$ (middle) and $A=0$ (right). In each panel of $R=10^2$,
$\kappa$ is taken to be 10$^{-4}$, 10$^{-3}$ and 10$^{-2}$. In the cases of
$R=10^4$, $\kappa$ is taken to be 10$^{-4}$, 10$^{-3}$.}
\end{figure}

The left panel of Figure 4 shows that the three curves of
$\kappa=10^{-4}$, 10$^{-3}$ and 10$^{-2}$ are almost the same. This
is consistent with Figure 2 that for Model I, the time-evolution of
$f$ are $\kappa$-independent for the pure scattering dust. For the pure
absorption dust (the right panel of Figure 4), the saturated flux is smaller
for larger $\kappa$. We can also see from Figure 4 that the time scale
of approaching saturation is smaller for larger $\kappa$. The result
of model II is in between that for models I and III.

With the saturated flux of Figure 4, one can define the effective
absorption optical depth by $\tau_{\rm effect}\equiv
-(1/\kappa)\ln f_{S}$. The results are shown in Table 1, in which $\tau_a$ is the
dust absorption depth. It is interested to see that the
effective absorption optical depth is always equal to about a few times of the
optical depth of resonant scattering $\tau_0$, regardless whether $\tau_a$ is less than 1.
Namely, the effective absorption depth $\tau_{\rm effect}$ of dust is roughly
proportional to $\tau_0$.

\begin{table}[htb]
\center{Table 1. Effective absorption optical depth $\tau_{\rm effect}$}
\begin{center}
\begin{tabular}{c|c|c|c|c|c|c|c}\hline
\multicolumn{2}{c|}{\rm }     &  \multicolumn{3}{c|}{\rm Model\ II} &
\multicolumn{3}{c}{\rm  Model\ III} \\ \hline
 $R=\tau_0$ & $\kappa$ & $\tau_a$ &  $f_S$   &   $\tau_{\rm effect}$  &  $\tau_a$   & $f_S$ &
   $\tau_{\rm effect}$ \\  \hline
$10^2$ & $10^{-4}$ & 0.0068 & 0.978  & $2.2\times 10^{2}$  & 0.01 & 0.963 & $3.8\times 10^{2}$\\
$10^2$ & $10^{-3}$ & 0.068  & 0.760  & $2.7\times 10^{2}$  & 0.10 & 0.670 & $4.0\times 10^{2}$  \\
$10^2$ & $10^{-2}$ & 0.68   & 0.116 & $2.2\times 10^2$  & 1.00 & 0.057 & $2.9\times 10^2$ \\
$10^4$ & $10^{-4}$ & 0.68   & $6.28\times 10^{-2}$ & $2.8\times 10^4$ & 1.00 & $3.02\times 10^{-2}$ &
        $3.5\times 10^4$ \\
$10^4$ & $10^{-3}$ & 6.8  &  $4.07\times 10^{-7}$ & $1.5\times 10^4$ & 10.0 &
    $2.87\times 10^{-9}$ &  $1.97 \times 10^4$ \\ \hline
\end{tabular}
\end{center}
\end{table}

According to the random walk scenario, if a medium has optical depths of
absorption $\tau_a$ and scattering $\tau_s$, the effective absorption optical depth
should be equal to $\tau_{\rm effect}=\sqrt{\tau_a(\tau_a+\tau_s)}$
(Rybicki \& Lightman 1979). However, the results of the last line of Table 1 show
that the random walk scenario does not work for the dust effect on resonant photon transfer.
This result is consistent with Figures 2 and 3. When optical depth of dust is lower than
the optical depth of resonant scattering $\tau_0$, the time scale of photon
escaping basically is not affected by the dust, but is proportional to $\tau_0$, and
therefore, the absorption is also proportional to $\tau_0$.

\subsection{Escape coefficient}

With the total flux, we can define the escaping coefficient of
Ly$\alpha$ photon as $f_{\rm esc}(\eta, \tau_0) \equiv F(\eta)/F_0$,
where $F_0$ is the flux of the center source. Figure 5 shows $f_{\rm
esc}(\eta, \tau_0)$ at three times $\eta=5\times 10^3$, 10$^4$ and
3.2$\times 10^4$ for Model II and $\kappa=10^{-3}$. At $\eta=5\times
10^3$, the flux of halos with $\tau_0 \leq 10^3$ is saturated. At
$\eta=10^4$, halos with $\tau_0\leq 3\times 10^3$ are saturated, and
all halos of $\tau_0 \leq 10^4$ are saturated at $\eta=3.2\times
10^4$.

\begin{figure}[htb]
\begin{center}
\includegraphics[scale=0.28]{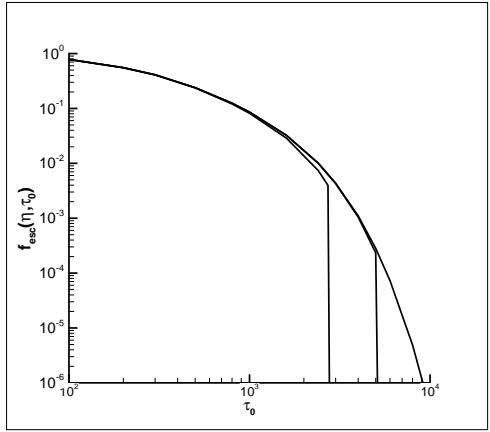}
\end{center}
\caption{Escaping coefficient $f_{\rm esc}(\eta)$ as a function of the
optical depth $\tau_0$ of halo at time $\eta = 5\times 10^3$, 10$^4$, and
3.2$\times 10^4$ from bottom to up. Dust is modeled by II, $A=0.32, \ g=0.73$,
and $\kappa=10^{-3}$. }
\end{figure}

\section{Dust effects on double-peaked profile}

\subsection{Dust and the frequency of double peaks }

A remarkable feature of Ly$\alpha$ photon emergent from optically thick medium is
the double-peaked profile. Figures 1, 2 and 3 have shown that the double
peak frequencies $x_{+}=|x_{-}|$ are almost independent of either the scattering
or the absorption of dust. In this section, we consider halos with size $R$ or $\tau_0$
larger than $10^2$. Figure 6 presents the double peak frequency $|x_{\pm}|$
as a function of $a\tau_0$, where the parameter $a$ is taken to be 10$^{-2}$ (left) and
$5\times 10^{-3}$ (right). Comparing the curves with dust and  without dust in Figure 6
we can say that the dust effect on $|x_{\pm}|$ is very small till $a\tau_0=aR=10^2$.

\begin{figure}[htb]
\begin{center}
\includegraphics[scale=0.25]{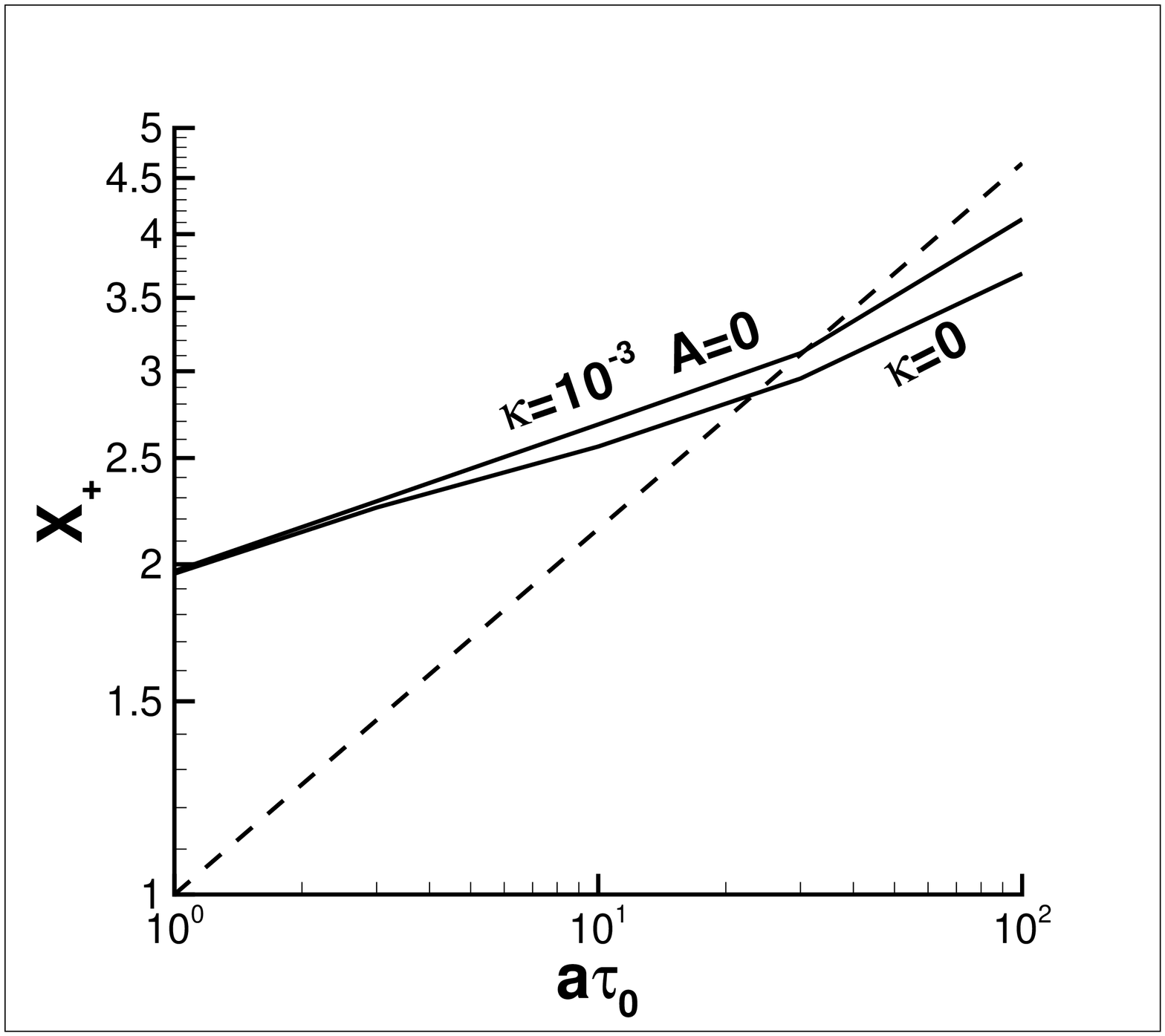}
\includegraphics[scale=0.25]{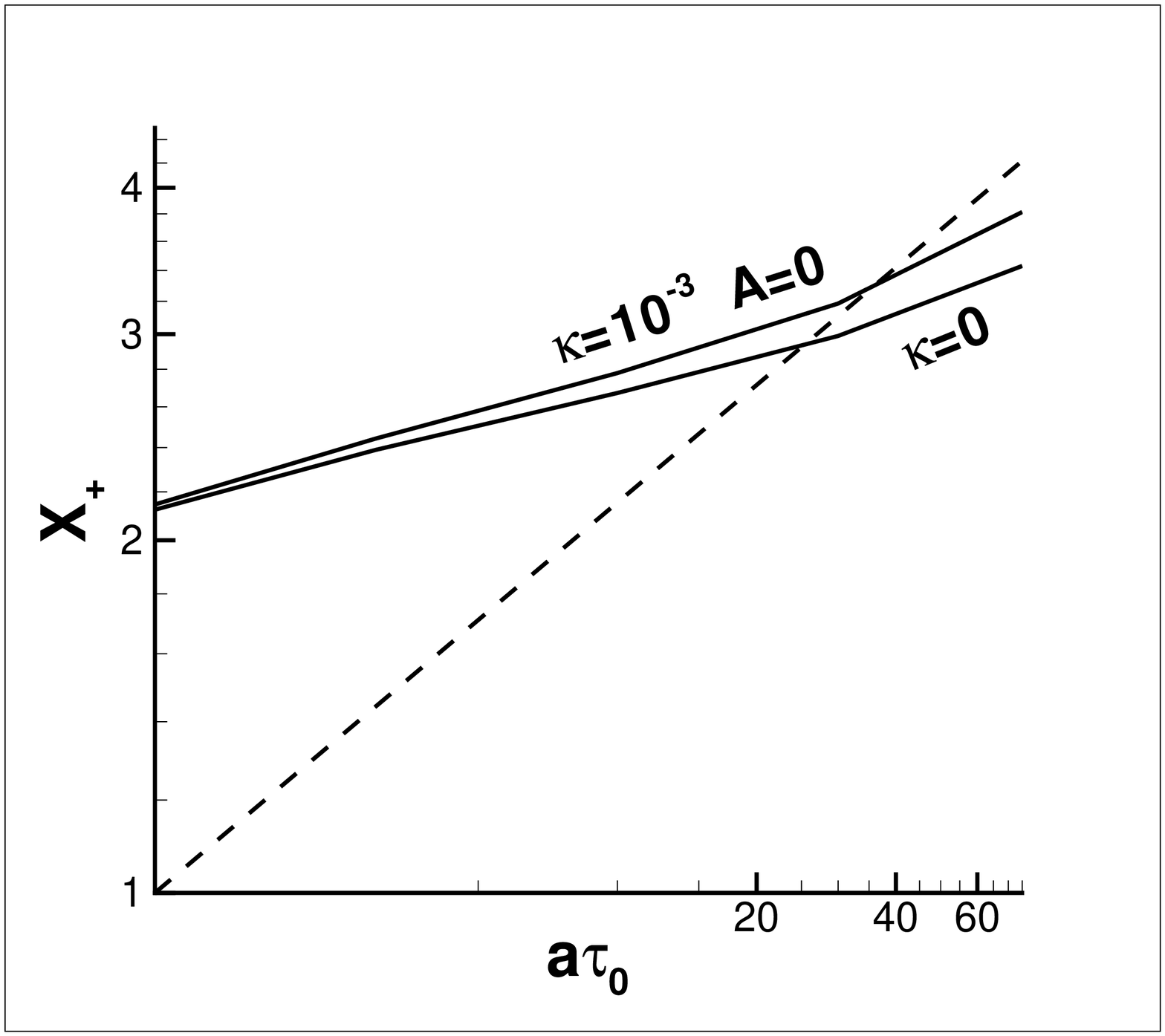}
\end{center}
\caption{The two-peak frequencies $x_{+}=|x_{-}|$ as a function of
$a\tau_0$. The parameter $a$ is taken to be 10$^{-2}$ (left) and
$5\times 10^{-3}$ (right). Dust model III (pure absorption) is used,
and $\kappa$ is taken to be $10^{-3}$. The dashed straight line
gives $\log x_{\pm}$-$\log a\tau$ with slope 1/3, which is to show
the $(a\tau)^{1/3}$-law of  $x_{\pm}$.}
\end{figure}

In the range $a\tau_0<20$, the $|x_{\pm}|$-$\tau_0$ relation
is almost flat with $|x_{\pm}|\simeq 2$. It is because the double-peaked
profile is given by the frequency range of the locally thermal
equilibrium. The positions of the two peaks, $x_{+}$ and $x_{-}$, basically are
at the maximum and minimum frequencies of the local thermalization. The frequency
range of the local thermal equilibrium state is mainly determined by the Doppler
broadening, and weakly dependent on $\tau_0$. Thus, we always have
$x_{\pm}\simeq \pm 2$. When the optical depth is larger, $a\tau_0\sim  10^2$, more and more
photons of the flux are attributed to the resonant scattering by the Lorentzian wing
of the Voigt profile. In this phase, $|x_{\pm}|$ will increase
with $\tau_0$.

Figure 6 shows also a line $x_{\pm} = \pm (a \tau_0)^{1/3}$, which is
given by the analytical solution of the Fokker-Planck approximation, in which the Doppler
broadening core in the Voigt profile
is ignored (Harrington 1973, Neufeld 1990, Dijkstra 2006). The numerical solutions of
eqs (3) or (9) and (10) deviate from the $(a\tau_0)^{1/3}$-law at all parameter range of
Figure 6. The deviation at $a\tau_0<20$ is due
to that the Doppler broadening core in the Voigt profile is ignored in the Fokker-Planck
approximation, and then, no locally thermal equilibrium can be reached.
Therefore, in the range $a\tau_0< 20$, $|x_{\pm}|$
of the WENO solution is larger than the  $(a\tau_0)^{1/3}$-law.
In the range of $a\tau_0>20$, the Fokker-Planck approximation yields
a faster diffusion of photons in the frequency space. This point
can be seen in the comparison between a Fokker-Planck solution with Field's
analytical solution (Figure 1 in Rybicki \& Dell'Antonio 1994). In this range, the numerical
results of $|x_{\pm}|$ is less than the $(a\tau_0)^{1/3}$-law.

\subsection{No narrowing and no widening}

The dust effect has been used to explain the narrowing of the width
between the two peaks (Laursen et al. 2009). Oppositely, it is also used
to explain the widening of the width between the two peaks (Verhamme et al. 2006). However,
Figures 1, 2, 3 and 6 already show that the width between the two peaks of the profile
is very weakly dependent on dust scattering and  absorption. This result supports, at
least in the parameter range considered in Figures 1, 2, 3, neither the narrowing nor the widening of
the two peaks.

\begin{figure}[htb]
\begin{center}
\includegraphics[scale=0.20]{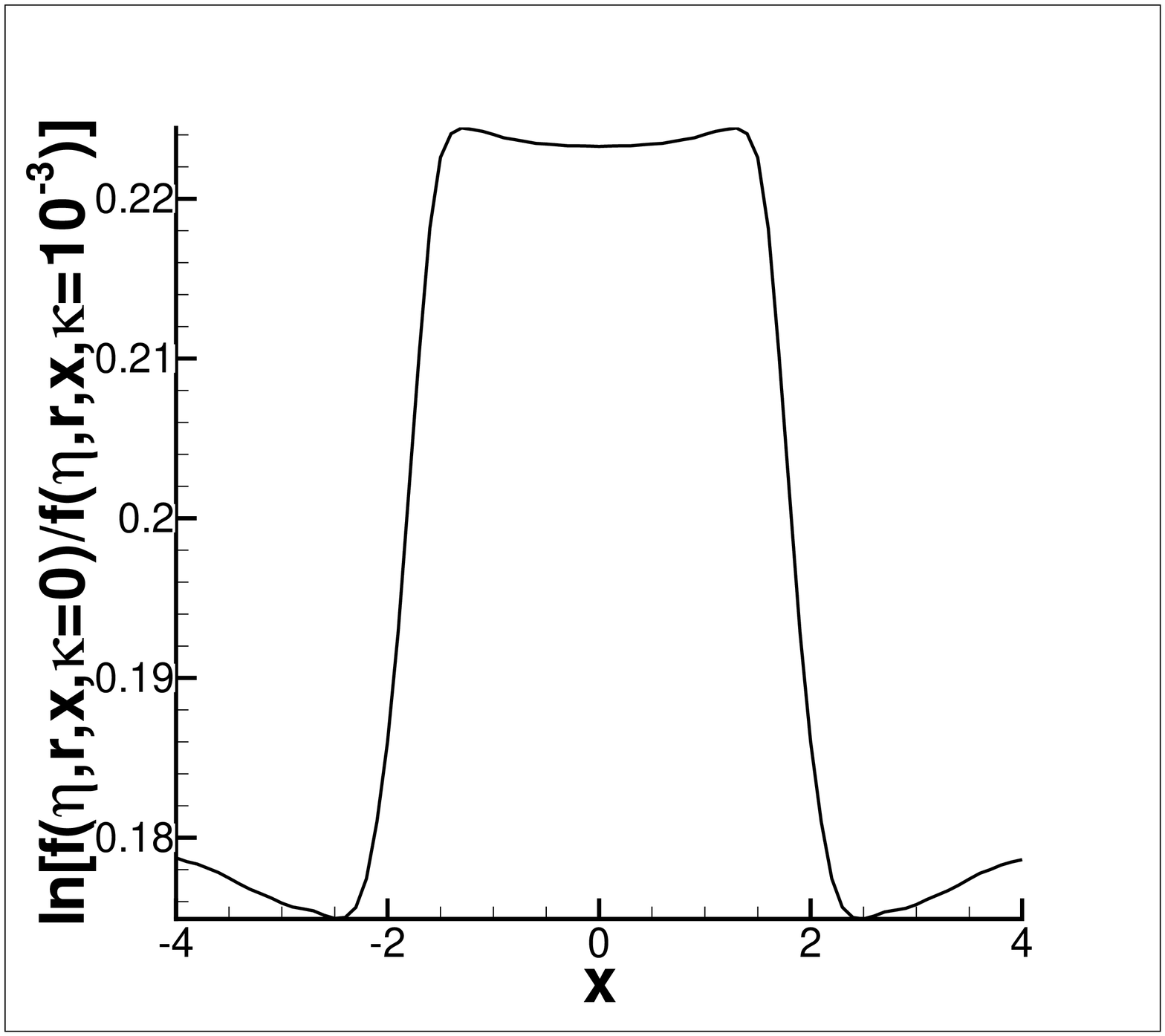}
\includegraphics[scale=0.20]{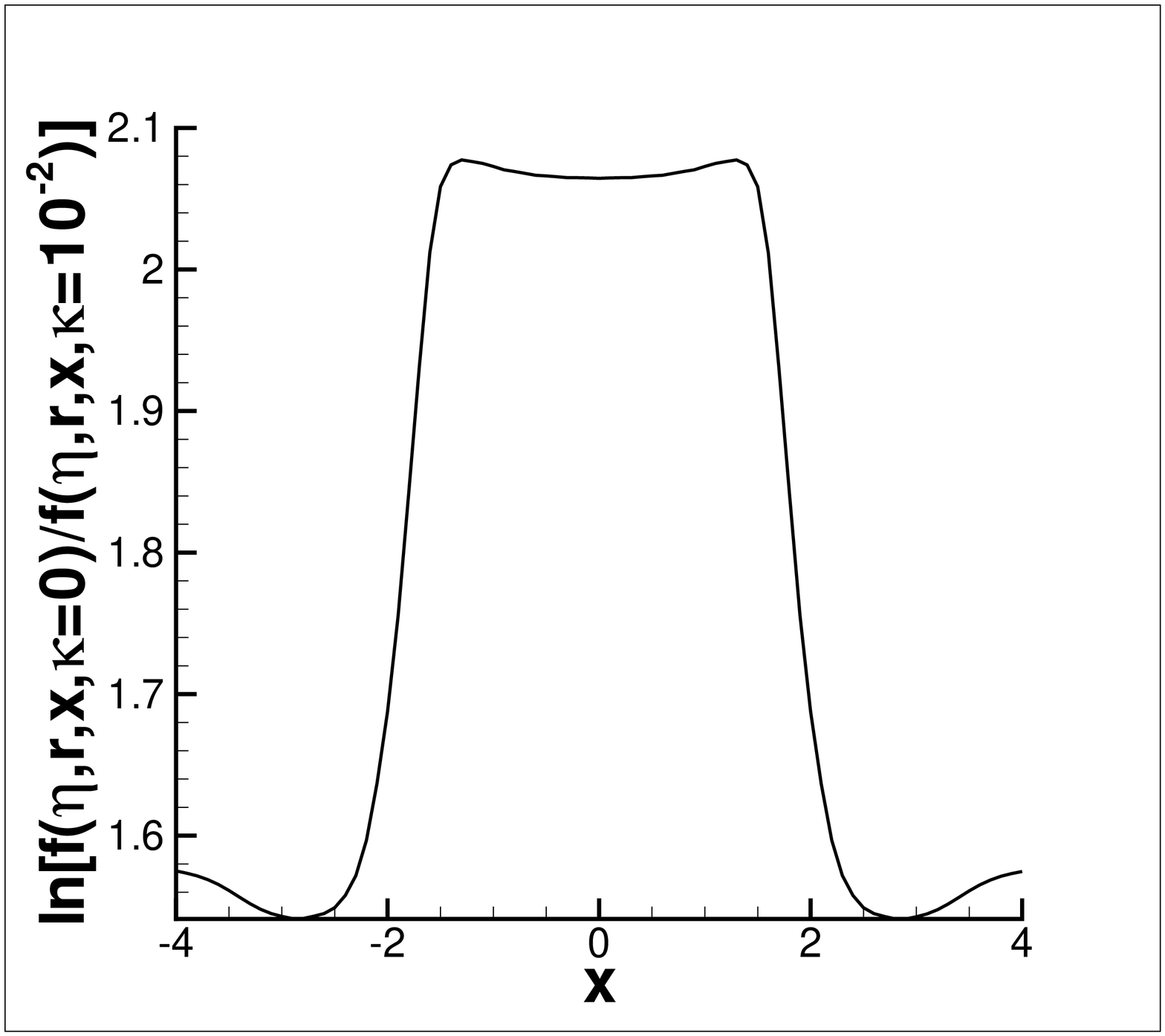}
\end{center}
\begin{center}
\includegraphics[scale=0.20]{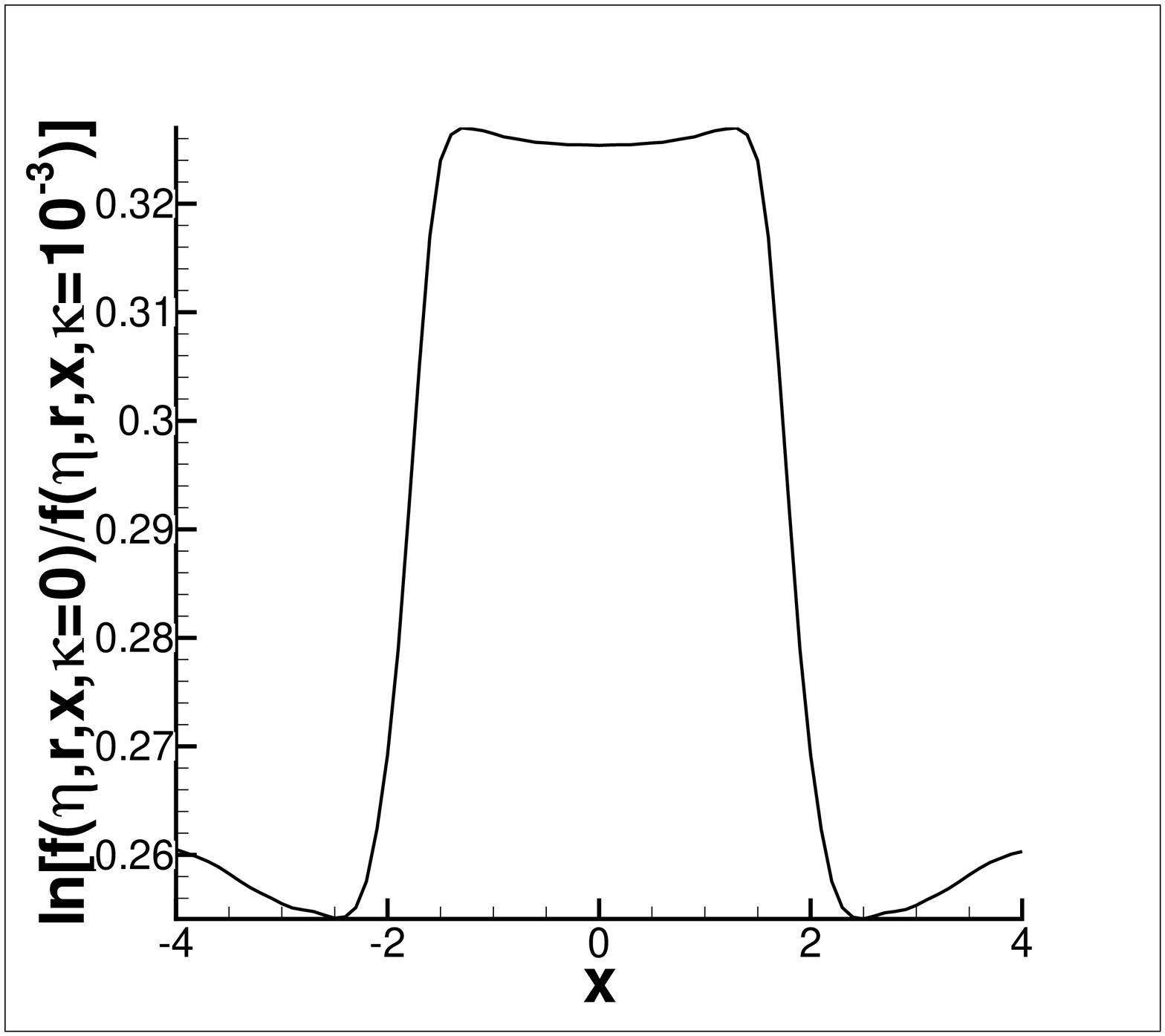}
\includegraphics[scale=0.20]{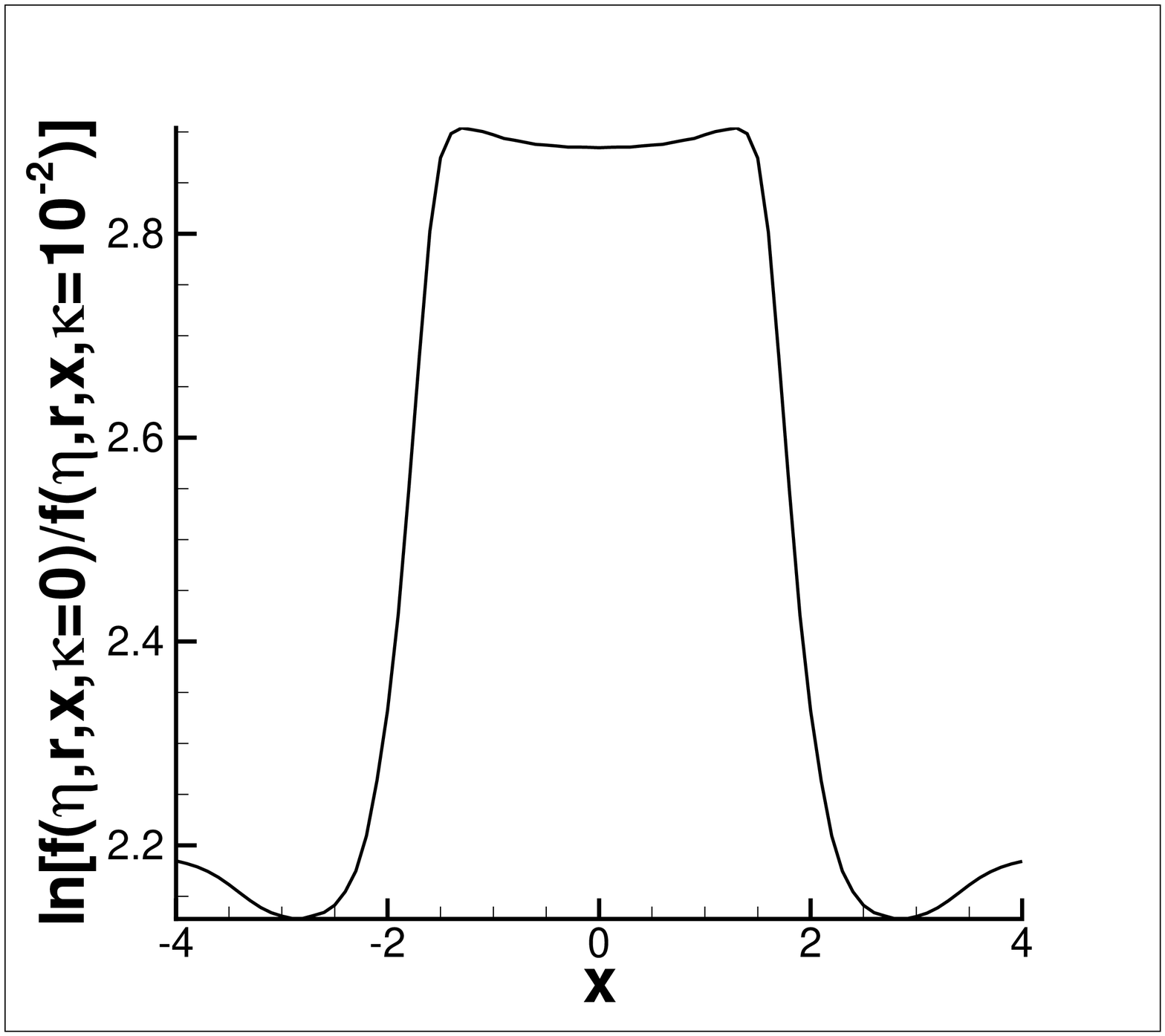}

\end{center}
\caption{$\ln [f(\eta, r,x,\kappa=0)/f(\eta, r,x,\kappa)]$ as function of
$x$ for model II
(up), and III (bottom), and $\kappa=10^{-3}$ (left) and 10$^{-2}$ (right). Other
parameters are the same as in Figure 2.
}
\end{figure}

If dust absorption can cause narrowing, the absorption should be weaker
at $|x|\sim 0$, and
stronger at $|x|\geq 2$. Similarly, if dust absorption can cause widening,
the absorption should be weaker at $|x|\sim 2$, and stronger at $|x|\sim 0$.
To test these assumptions,
Figure 7 plots $\ln [f(\eta, r,x,\kappa=0)/f(\eta, r,x,\kappa)]$ as
a function of $x$. It
measures the $x$(frequency)-dependence of the flux ratio with and
without dust absorption. We take large $\eta$, and then
the fluxes in Figure 7 are saturated. Figure 7 shows that the absorption in the range
$|x|<2$ is much stronger than that of $|x|>2$, and therefore, the assumption of the
narrowing is ruled out. Figure 7 shows also that the curves of
$\ln [f(\eta, r,x,\kappa=0)/f(\eta, r,x,\kappa=10^{-3})]$ are almost flat in the range
$|x|<2$. Therefore, the assumption of widening of the two peaks can also be ruled out.

Since the cross sections of dust absorption and scattering are assumed
to be frequency-independent.
Eqs. (9) and (10) do not contain any frequency scales other than that from
resonant scattering.
However, either narrowing or widening would require to have frequency scales
different from that of
resonant scattering. This is occurence is not possible if the dust is gray.

\subsection{Profile of absorption spectrum}

If the radiation from the sources has a continuum spectrum, the
effect of neutral hydrogen halos is to produce an absorption line at
$\nu=\nu_0$. The profile of the absorption line can also be found by
solving equations (9) and (10), but replacing the boundary equation
(11) by
\begin{equation}
j(\eta, 0, x)=0, \hspace{1cm} f(\eta, 0, x)=S_0.
\end{equation}
That is, we assume that the original spectrum is flat in the
frequency space. The spectrum of the flux emergent from halo of $R=10^2$ and $10^4$ with
central source of eq.(14) for dust models I, II and III are shown in Figure 8.
All curves are for large $\eta$, i.e. they are saturated.

\begin{figure}[htb]
\begin{center}
\includegraphics[scale=0.20]{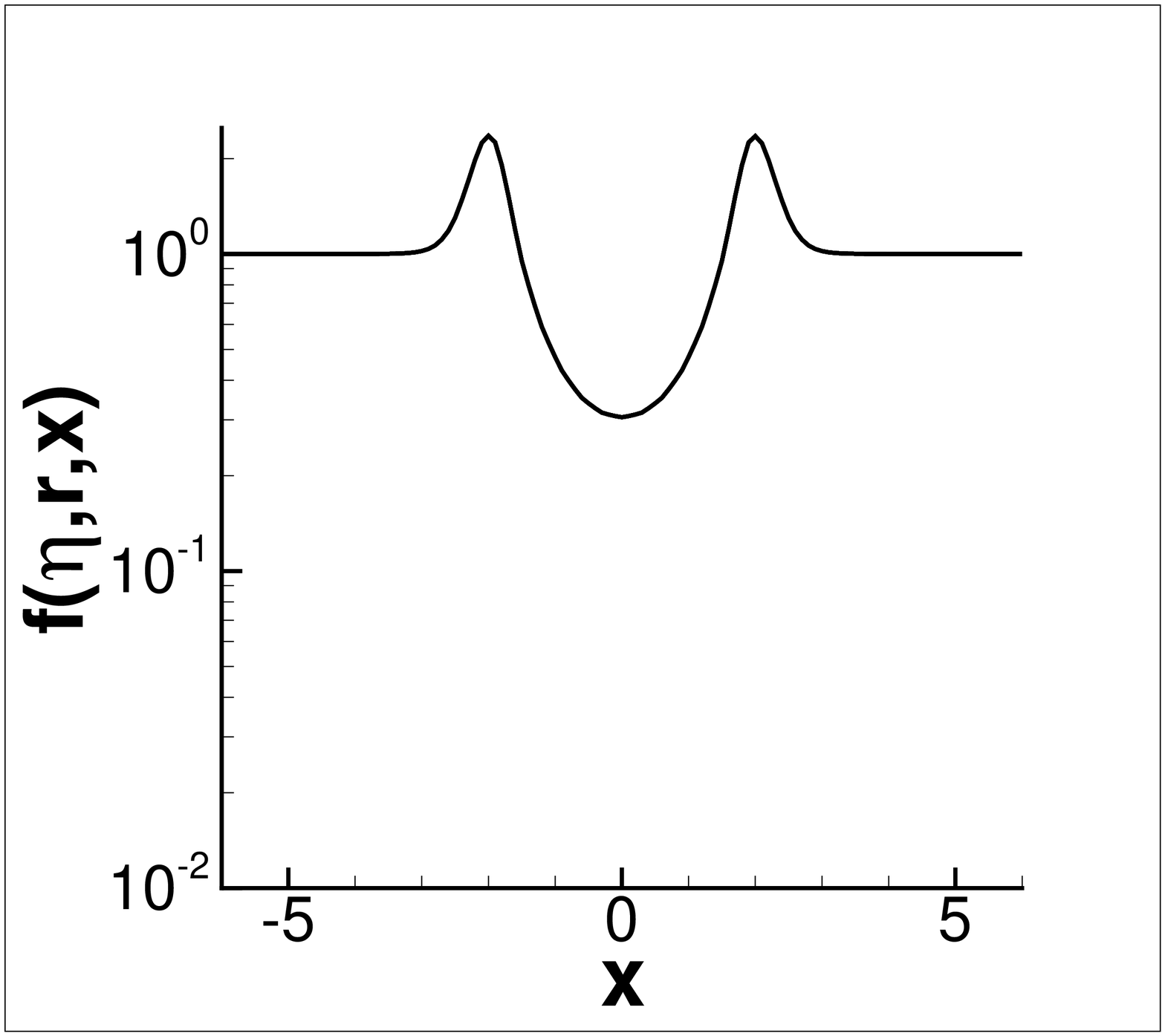}
\includegraphics[scale=0.20]{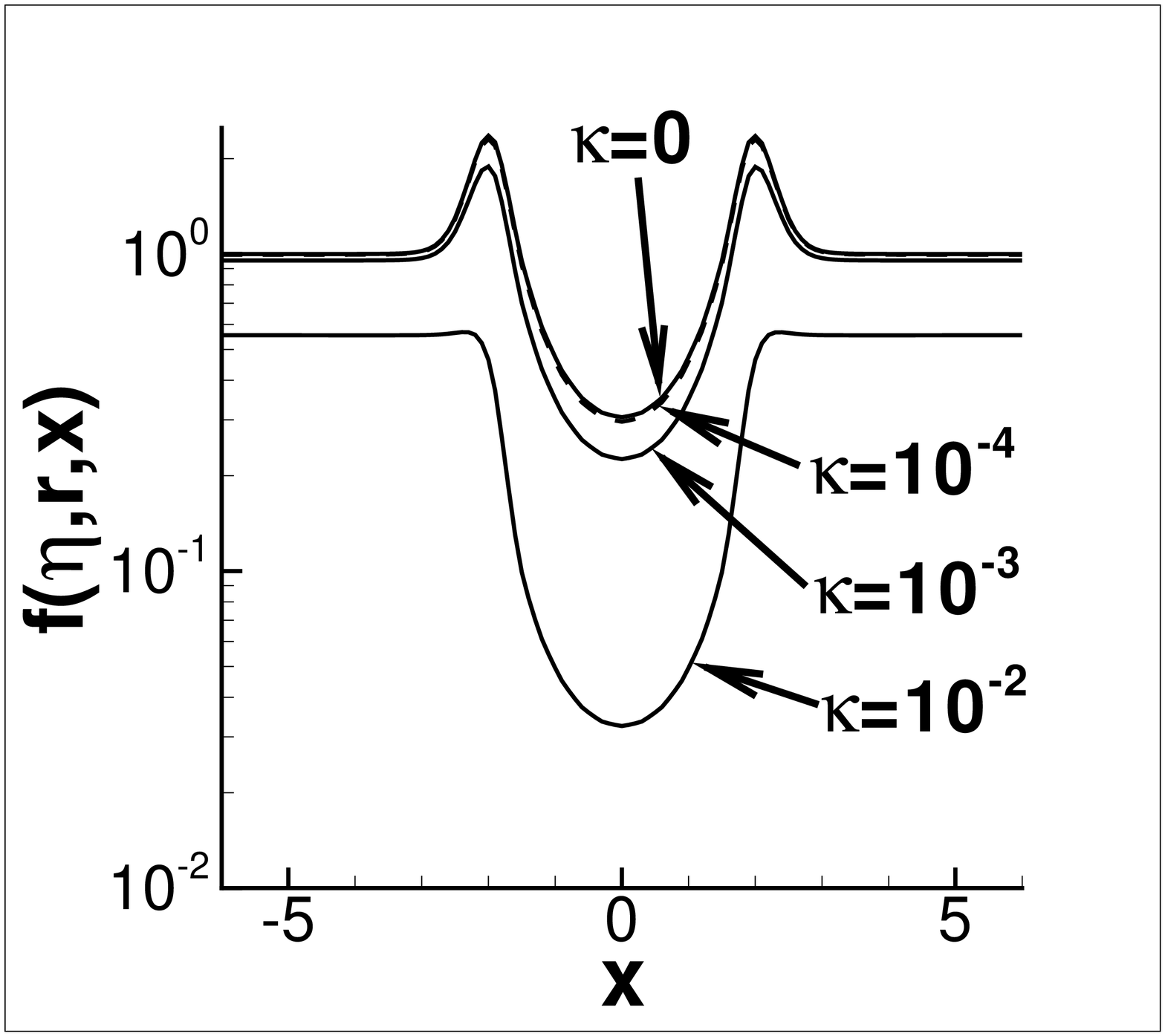}
\includegraphics[scale=0.20]{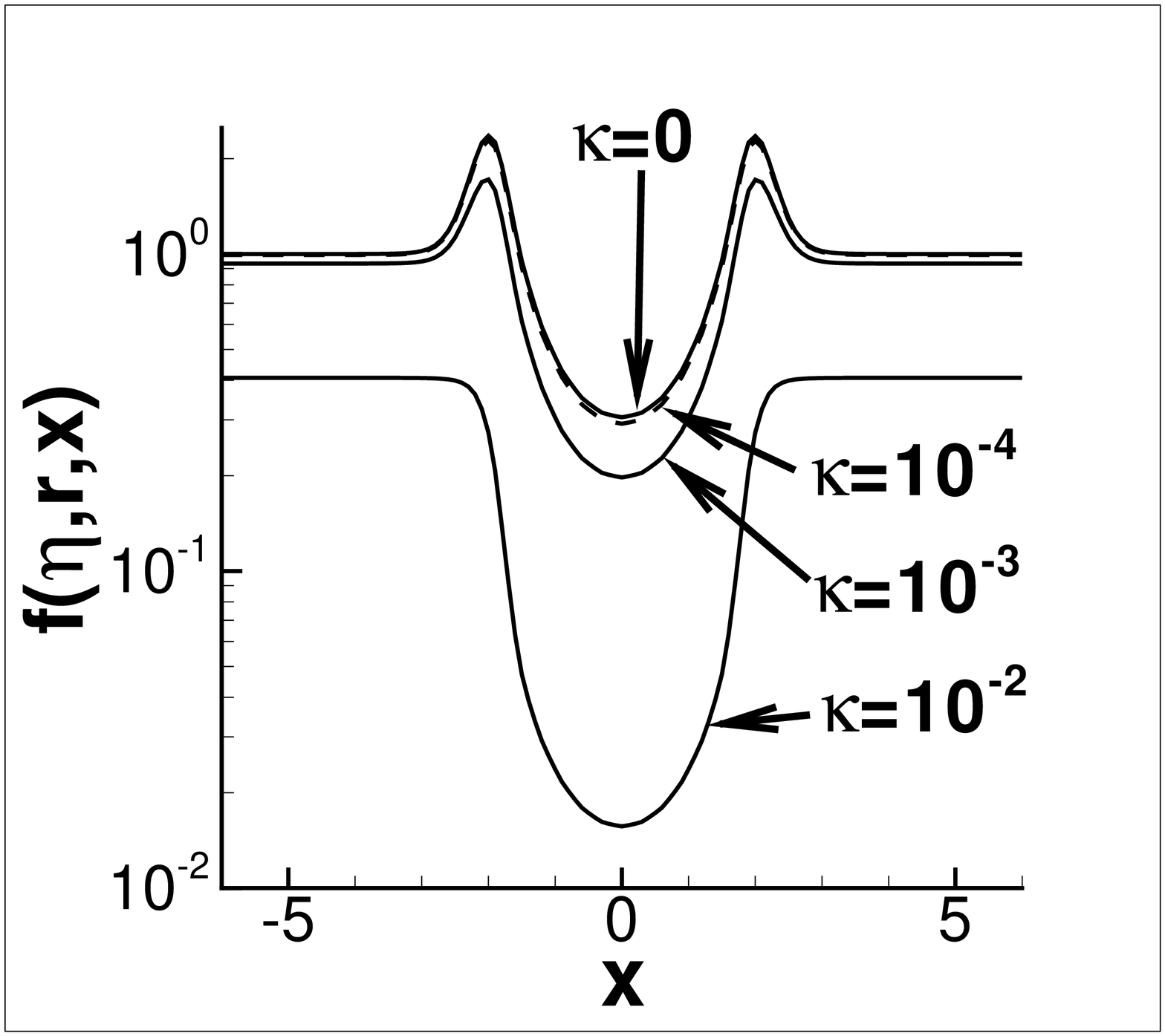}
\end{center}
\begin{center}
\includegraphics[scale=0.20]{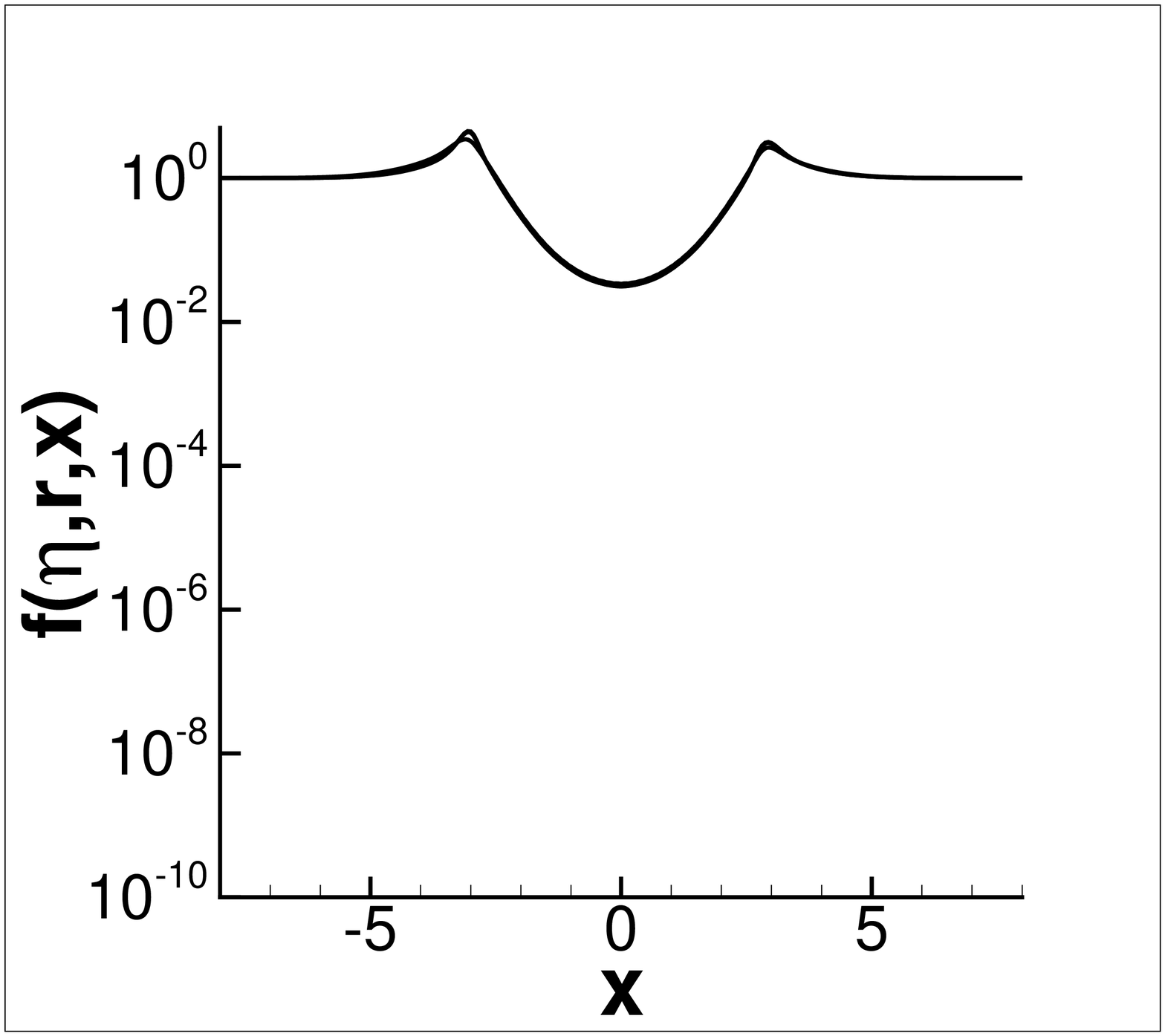}
\includegraphics[scale=0.20]{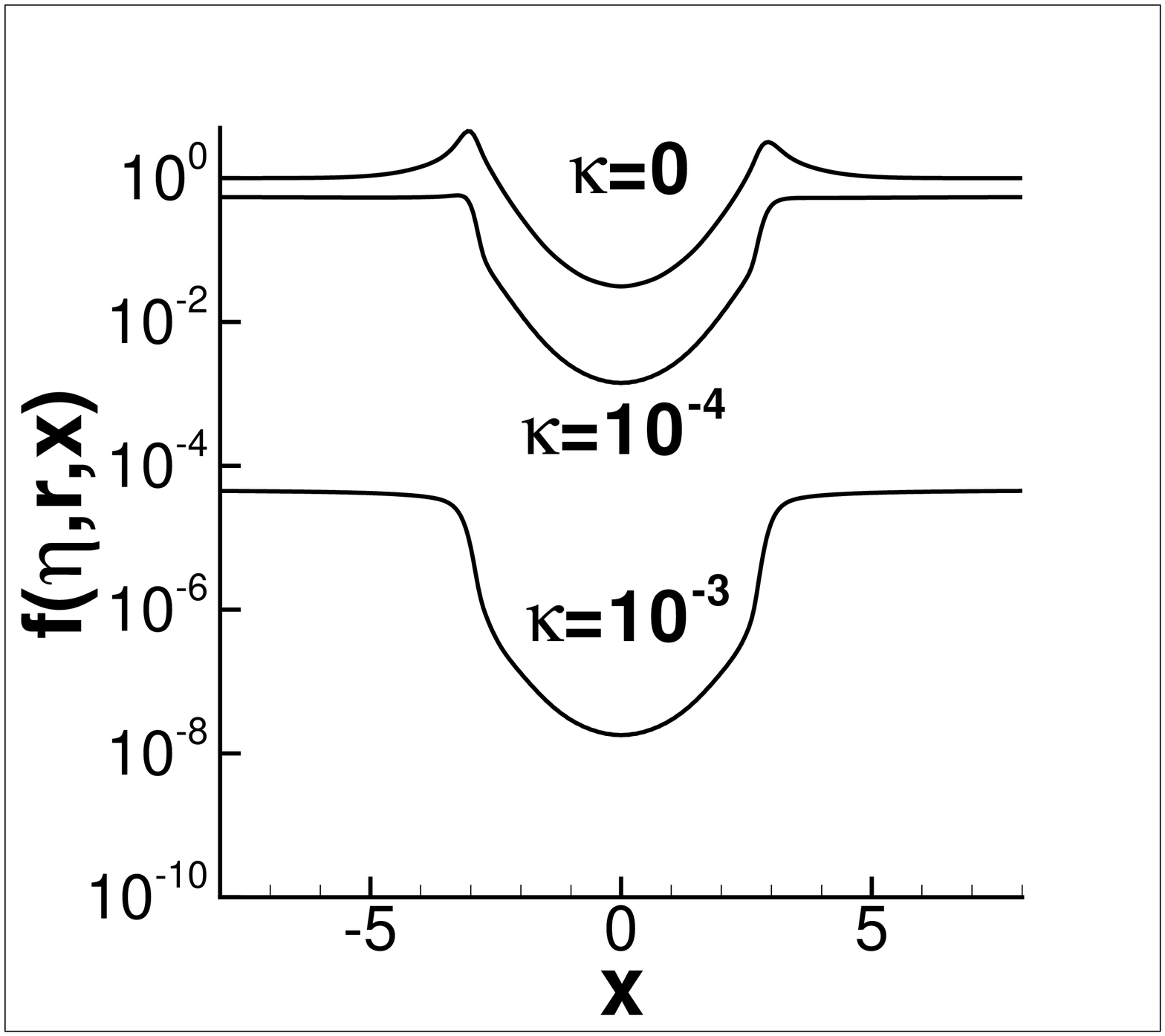}
\includegraphics[scale=0.20]{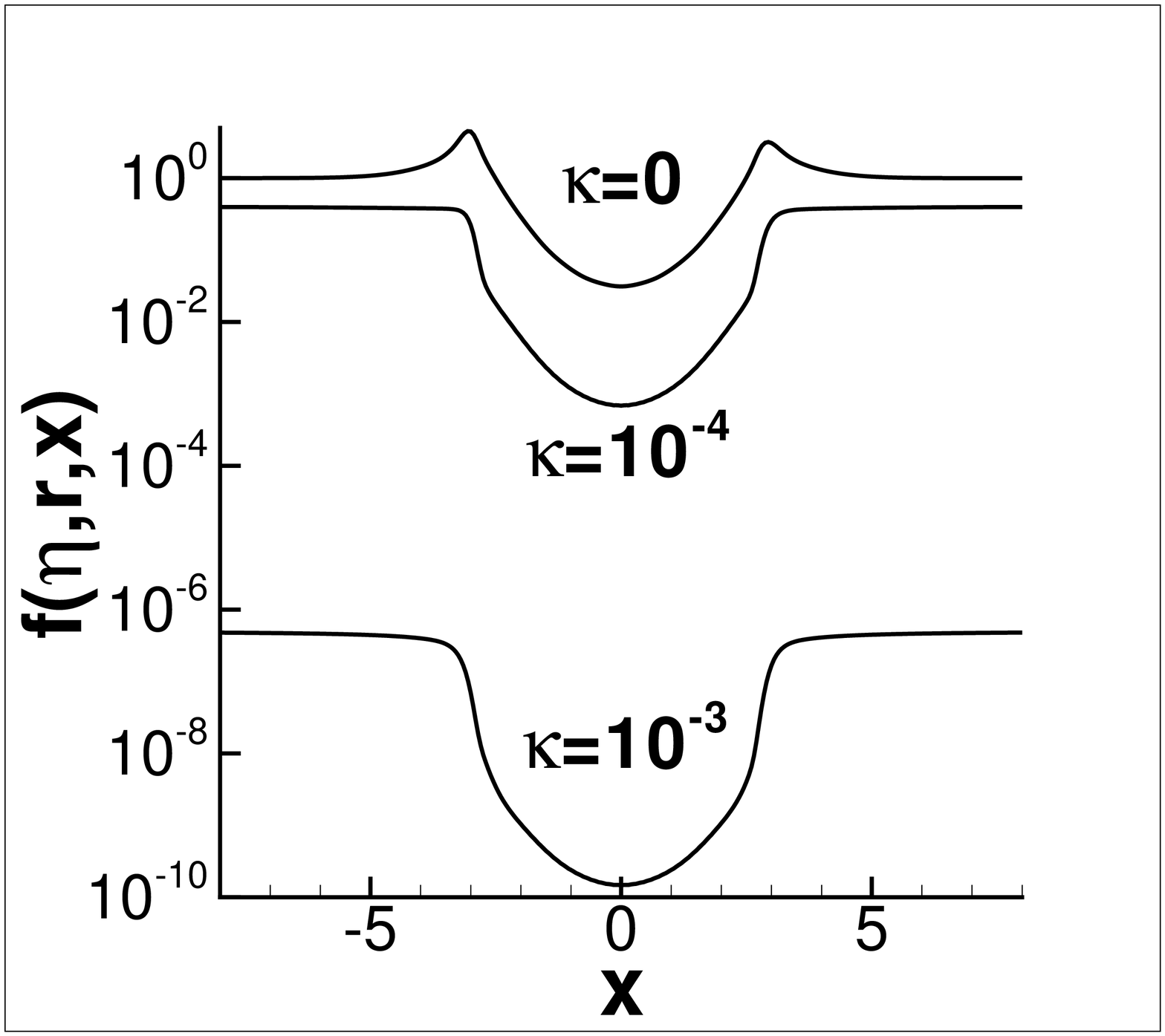}
\end{center}
\caption{The spectrum of the flux emergent from halo of $R=10^2$ (upper panels) and
10$^4$ (lower panels)
with central source of eq.(14) for the dust model I (left), II
(middle) and III (right). Other parameters are the same as
in Figure 2.}
\end{figure}

The optical depths at the frequency $|x| >4$ are small, and therefore, the
Eddington approximation might no longer be proper. However, those photons do
not strongly involve the resonant scattering, and hence they do not
significantly affect the solution around $x=0$. The
solutions of Figure 8 is still useful to study the profiles of $f$ around $x=0$.

The flux profile of Figure 8 typically are absorption lines with width
given by the double peaks similar to the double peaked structure of the emission line.
The flux at the double peaks is even higher than the flat wing. It is because more
photons are stored in the frequency
range $|x|<2$. According to the redistribution function eq.(4), the probability
of transferring a $x'$ photon to a $|x|<|x'|$ photon is larger than that from
$|x'|$ to $|x|>|x'|$. Therefore, if the original spectrum is flat, the net effect
of resonant scattering is to bring photons with frequency $|x|>2$ to $|x|<2$.
Photons stored $|x|<2$ are thermalized, and therefore, in the range $|x|<2$,
the profile will be
the same as the emission line, and the double peaks can be higher than the wing.
It makes the shoulder at $|x|\sim 2$.

As expected, for model I (left panels of Figure 8), the double profile is completely
$\kappa$-independent. Dusty scattering does not change the flux and its profile. For
models II and III, the higher the $\kappa$, the
lower the flux of the wing, because the dust absorption is assumed to be
frequency-independent. The positions of the double peaks, $x_{\rm}$, in the absorption
spectrum are also $\kappa$-independent. This once again shows that dust
absorption and scattering causes neither narrowing nor widening of the
double-peaked
profile. However, for higher $\kappa$ the flux of the peaks is lower.
When the absorption is very strong, the double-peaked structure might disappear, but
will never be narrowed or widened.

\section{Discussions and conclusions}

The study of dust effects on radiative transfer has had a long history
related to extinction. However, dust effects on radiative transfer of
resonant photons actually have not been carefully investigated. Existing
works are mostly based on the solutions of the Fokker-Planck
approximation, or Monte Carlo simulation. These results are important.
We revisited these problems with the WENO solver of the integro-differential
equation of the resonant radiative transfer, and have found some features which
have not been addressed in previous works.  These features are
summarized as follows.

First, the random walk picture in the physical space will no longer be
available for estimating the effective optical depth of dust absorption. For a medium
with the optical depth of absorption and resonant scattering to be $\tau_a\gg 1$, $\tau(\nu_0)\gg 1$
and $\tau_s(\nu_0)\gg \tau_a$, the effective absorption optical depth is found to be
almost independent of $\tau_a$, and to be equal to about a few times of $\tau_s(\nu_0)$.

Second, dust absorption will, of course, yield the decrease of the flux of Ly$\alpha$
photons emergent from optical thick medium. However, if the absorption
cross-section of dust is frequency independent, the double-peaked structure of the
frequency profile is basically dust-independent. The double-peaked structure does not get
narrowed or widened by the absorption and scattering of dust.

Third, the time scales of Ly$\alpha$ photon transfer basically are
independent of dust scattering and absorption. It is because those time scales
are mainly determined by the kinetics in the frequency space, while dust
does not affect the behavior of the transfer in the frequency space if the cross
section of the dust is wavelength-independent. The local thermal equilibrium makes
the anisotropic scattering to be ineffective on the angular distribution of photons. Dust
absorption and scattering do not lead to the increase or decrease of the time of
storing Ly$\alpha$ photons in the halos.

The differences between the time-independent solutions of the Fokker-Planck
approximation, or Monte Carlo simulation and the WENO solution of eq.(3) is mainly related
to the W-F effect. Therefore, all above-mentioned features can already be clearly seen with
halos of $\tau_0 \sim 10^2$, in which the W-F local thermal equilibrium has been well
established.

In this context, most calculation in this paper is on holes with $\tau_0<10^5$. This range of
$\tau_0$ certainly is unable to describe halos with column number density of HI larger than
10$^{17}$ cm$^{-2}$ (e.g. Roy et al. 2010). Nevertheless, the result of $\tau_0<10^5$ would already
be useful  for studying the 21 cm region around high-redshift
sources, of which the optical depth typically is (Liu et al 2007; Roy, et al. 2009c).
\begin{equation}
\tau_0=3.9\times 10^5 f_{\rm HI}\left (\frac{T}{10^4 {\rm K}}\right )^{-1/2}
\left(\frac{1+z}{10}\right)^{3}\left ( \frac{\Omega_bh^2}{0.022}\right )
\left ( \frac{R_{\rm ph}}{10 {\rm kpc}}\right ),
\end{equation}
where $f_{\rm HI}$ is the fraction of HI. All other parameters in eq. (15) is taken from the concordance
$\Lambda$CDM mode. For these objects the relation between dimensionless $\eta$ and physical time $t$ is given by
\begin{equation}
t=5.4 \times 10^{-2}f^{-1}_{\rm HI} \left (\frac{T}{10^4 {\rm K}}\right )^{1/2}
\left(\frac{1+z}{10}\right)^{-3}\left ( \frac{\Omega_bh^2}{0.022}\right )^{-1} \eta, \ \ {\rm yr}.
\end{equation}
The 21 cm emission rely on the W-F effect. On the other hand, the time-scale of the evolution of the
21 region is short. The effect of dust on the time-scales of Ly$\alpha$ evolution should be considered.

We have not considered the Ly$\alpha$ photons produced by the recombination in
the ionized halo. If the halo is optical thick, photons from the
recombination will also be thermalized. The information of where the photon
comes from will be forgotten during the thermalization. Therefore, photons from
recombination should not show any difference from those emitted from central sources.
Only the photons formed at the place very close to the boundary of the halo will not
be thermalized, and may yield different behavior.

\acknowledgments

This research is partially supported by ARO grants
W911NF-08-1-0520 and W911NF-11-1-0091.

\appendix

\section{Integral of the phase function [eq.(6)]}

Eq.(6) can be rewritten as
\begin{equation}
\mathcal{R}^d(\mu,\mu')=\frac{1}{4\pi}\int^{2\pi}_{0}
   d\phi' \frac{1-g^2}{|{\bf I}-g{\bf I'}|^{\frac{3}{2}}}
\end{equation}
where ${\bf I}$ and ${\bf I'}$ are unit vector on the direction of
polar angle $\theta$ and $\theta'$, and azimuth angle $\phi$ and $\phi'$,
respectively. That is ${\bf I}\cdot{\bf I}={\bf I'}\cdot{\bf I'}=1$
and ${\bf I}\cdot{\bf I'}=\cos\gamma= \cos\theta \cos\theta' +
\sin \theta\sin\theta' \cos(\phi-\phi')$, and $\mu=\cos\theta$, $\mu'=\cos \theta$. We have
\begin{eqnarray}
\frac{d}{dg}\frac{1}{|{\bf I}-g{\bf I'}|^{1/2}}
= \frac{1- g^2}{2g|{\bf I}-g{\bf I'}|^{3/2}}-\frac{1}{2g|{\bf I}-g{\bf I'}|^{1/2}},\\ \nonumber
\end{eqnarray}
and therefore, 
\begin{equation}
\frac{1- g^2}{|{\bf I}-g{\bf I'}|^{3/2}}=2g\frac{d}{dg}\frac{1}{|{\bf I}-g{\bf I'}|^{1/2}}+
\frac{1}{|{\bf I}-g{\bf I'}|^{1/2}} .
\end{equation}

The expansion with Legendre functions $P_l(\cos\gamma)$ gives
\begin{equation}
\frac{1}{|{\bf I}-g{\bf I'}|^{1/2}}=\sum_{l=0}^{\infty} g^lP_l(\cos\gamma),
\end{equation}
and then
\begin{equation}
\frac{1- g^2}{|{\bf I}-g{\bf I'}|^{3/2}}=\sum_{l=1}^{\infty}2 l g^lP_l(\cos\gamma)+
\sum_{l=0}^{\infty} g^lP_l(\cos\gamma).
\end{equation}

Since $\cos\gamma= \cos\theta \cos\theta' +
\sin \theta\sin\theta' \cos(\phi-\phi')$, we have the following identity
for the Legendre function $P_l(\cos\gamma)$ as
\begin{equation}
P_l(\cos \gamma)=P_l(\cos\theta)P_l(\cos\theta')+
2\sum_{m=1}^{m=l}\frac{(l-m)!}{(l+m)!}P^m_l(\cos\theta)
  P^m_l(\cos\theta') \cos [m(\phi-\phi')] .
\end{equation}

The integral of $\phi'$ in eq.(A1) kills the second term of eq.(A6), we have then
\begin{eqnarray}
\mathcal{R}^d(\mu,\mu') & =&\frac{1}{4\pi}2\pi
\left [\sum_{l=1}^{\infty}2 l g^l P_l(\cos\theta)P_l(\cos\theta')+
\sum_{l=0}^{\infty} g^lP_l(\cos\theta)P_l(\cos\theta')\right ] \\ \nonumber
& = & \frac{1}{2}\left [ \sum_{l=1}^{\infty}2 l g^l P_l(\mu)P_l(\mu')+
\sum_{l=0}^{\infty} g^lP_l(\mu)P_l(\mu') \right ] . \\ \nonumber
\end{eqnarray}

Using the orthogonal relation $\int_{-1}^{1}P_l(\mu)P_{l'}(\mu)d\mu =
\frac{2}{2l+1}\delta_{l,l'}$,
we have
\begin{equation}
R_0(g)  = \frac{1}{2} \int_{-1}^{1}d\mu \int_{-1}^1 d\mu' R^d(\mu,\mu')=1,
\end{equation}
for which only the term $l=0$ in eq.(A7) has contribution. Similarly,
\begin{equation}
R_1(g)  = \frac{1}{2}\int_{-1}^{1}d\mu \int_{-1}^1 d\mu' \mu R^d(\mu,\mu')=
 \frac{1}{2}\int_{-1}^{1}d\mu \int_{-1}^1 d\mu' \mu' R^d(\mu,\mu')= 0,
\end{equation}
\begin{equation}
R_2(g)  = \frac{1}{2} \int_{-1}^{1}d\mu \int_{-1}^1 d\mu' \mu \mu' R^d(\mu,\mu')
=\frac{g}{3}.
\end{equation}
These results are used in deriving eqs.(9) and (10).

\section{Numerical algorithm}

To solve Equations $(9)$ and $(10)$ as a system, our computational
domain is
$(r,x)\in[0,r_\textrm{max}]\times[x_\textrm{left},x_\textrm{right}],$
where $r_\textrm{max},x_\textrm{left}$ and $x_\textrm{right}$ are
chosen such that the solution vanishes to zero outside the
boundaries. We choose mesh sizes with grid refinement tests to
ensure proper numerical resolution. In the following, we describe
numerical techniques involved in our algorithm, including
approximations to spatial derivatives, numerical boundary condition,
and time evolution.
\subsection{The WENO Algorithm: Approximations to the Spacial Derivatives}
The spacial derivative terms in Equation $(9)$ and $(10)$ are
approximated by a fifth-order finite difference WENO scheme.
\par
We first give the WENO reconstruction procedure in approximating
$\frac{\partial j}{\partial x},$
\begin{displaymath}
\frac{\partial j(\eta^n,r_i,x_j)}{\partial x}\approx\frac{1}{\Delta
x}(\hat{h}_{j+1/2}-\hat{h}_{j-1/2})
\end{displaymath}
with fixed $\eta=\eta^n$ and $r=r_i.$ The numerical flux
$\hat{h}_{j+1/2}$ is obtained by the fifth-order WENO approximation
in an upwind fashion, because the wind direction is fixed. Denote
\begin{displaymath}
h_j=j(\eta^n,r_i,x_j),\quad\quad j=-2,-1,\cdots,N+3
\end{displaymath}
with fixed $n$ and $i$. The numerical flux from the WENO procedure
is obtained by
\begin{displaymath}
\hat{h}_{j+1/2}=\omega_1\hat{h}^{(1)}_{j+1/2}+\omega_2\hat{h}_{j+1/2}^{(2)}+\omega_3\hat{h}_{j+1/2}^{(3)}
\end{displaymath}
where $\hat{h}_{j+1/2}^{(m)}$ are the three third-order fluxes on
three different stencils given by
{\setlength\arraycolsep{2pt}
\begin{eqnarray*}
\hat{h}_{j+1/2}^{(1)}&=&-\frac{1}{6}h_{j-1}+\frac{5}{6}h_j+\frac{1}{3}h_{j+1},\\
\hat{h}_{j+1/2}^{(2)}&=&\frac{1}{3}h_j+\frac{5}{6}h_{j+1}-\frac{1}{6}h_{j+2},\\
\hat{h}_{j+1/2}^{(3)}&=&\frac{11}{6}h_{j+1}-\frac{7}{6}h_{j+2}+\frac{1}{3}h_{j+3}.
\end{eqnarray*}}
and the nonlinear weights $\omega_m$ are given by
\begin{displaymath}
\omega_m=\frac{\breve{\omega}_m}{\sum_{l=1}^3\breve{\omega}_l},
\quad\breve{\omega}_l=\frac{\gamma_l}{(\epsilon+\beta_l)^2}
\end{displaymath}
where $\epsilon$ is a parameter to avoid the denominator to become
zero and is taken as $\epsilon=10^{-8}$. The linear weights
$\gamma_l$ are given by
\begin{displaymath}
\gamma_1=\frac{3}{10},\quad\gamma_2=\frac{3}{5},\quad\gamma_3=\frac{1}{10},
\end{displaymath}
and the smoothness indicators $\beta_l$ are given by
{\setlength\arraycolsep{2pt}
\begin{eqnarray*}
\beta_1&=&\frac{13}{12}(h_{j-1}-2h_j+h_{j+1})^2+\frac{1}{4}(h_{j-1}-4h_j+3h_{j+1})^2,\\
\beta_2&=&\frac{13}{12}(h_j-2h_{j+1}+h_{j+2})^2+\frac{1}{4}(h_j-h_{j+2})^2,\\
\beta_3&=&\frac{13}{12}(h_{j+1}-2h_{j+2}+h_{j+3})^2+\frac{1}{4}(3h_{j+1}-4h_{j+2}+h_{j+3})^2.
\end{eqnarray*}}
\par
To approximate the $r$-derivatives in the system of Equations $(9)$
and $(10)$, we need to perform the WENO procedure based on
a characteristic decomposition. We write the left-hand side of
Equations $(9)$ and $(10)$ as
\begin{displaymath}
{\bf u}_t+A{\bf u}_r,
\end{displaymath}
where ${\bf u}=(j,f)^T$ and
\begin{displaymath}
A=\left(
\begin{array}{cc}
0&1\\
\frac{1}{3}&0
\end{array}
\right)
\end{displaymath}
is a constant matrix. To perform the characteristic decomposition,
we first compute the eigenvalues, the right eigenvectors and the
left eigenvectors of A and denote them by $\Lambda$, $R$ and
$R^{-1}$. We then project {\bf u} to the local characteristic fields
{\bf v} with ${\bf v}=R^{-1}{\bf u}.$ Now ${\bf u}_t+A{\bf u}_r$ of
the original system is decoupled as two independent equations as
${\bf v}_t+\Lambda{\bf v}_r.$ We approximate the derivative ${\bf
v}_r$ component by component, each with the correct upwind
direction, with the WENO reconstruction procedure similar to the
procedure described above for $\frac{\partial j}{\partial x}$. In
the end, we transform ${\bf v}_r$ back to the physical space by
${\bf u}_r=R{\bf v}_r$. We refer the readers to Cockburn et al. 1998
for more implementation details.
\subsection{Numerical Boundary Condition}
To implement the boundary condition $(12)$, we also need to perform
a characteristic decomposition as discussed above.  Using the same
notation as before, we project {\bf u}
to the local characteristic fields {\bf v} with ${\bf v}=R^{-1}{\bf
u}$.  Denote ${\bf v}=(v_1,v_2)^T$, now ${\bf u}_t+A{\bf u}_r$ of the
original system is decoupled to two independent scalar operators given by
$$
\frac{\partial v_1}{\partial t}+\lambda_1\frac{\partial
v_1}{\partial r}; \qquad
\frac{\partial v_2}{\partial t}+\lambda_2\frac{\partial
v_2}{\partial r}
$$
where $\lambda_1=\frac{\sqrt{3}}{3}$ and
$\lambda_2=-\frac{\sqrt{3}}{3}$. The characteristic line starting from
the boundary $r=r_{\textrm{max}}$ for the first equation is pointing
outside the computational domain while the one for the second
equation is pointing inside. For well-posedness of our system, we
need to impose the boundary condition there as
$$v_2=\alpha v_1+\beta$$
with constants $\alpha$ and $\beta$.
We can calculate the values of $\alpha$ and $\beta$ based on
equation $(12)$ and the left and right eigenvectors of $A$. For example,
if we take
$$R=\left(\begin{array}{cc}
\frac{\sqrt{3}}{2}&\frac{\sqrt{3}}{2}\\
\frac{1}{2}&-\frac{1}{2}
\end{array}\right),$$
we can compute that $\alpha=7-4\sqrt{3}$ and $\beta=0$. We use
extrapolation to obtain the value of $v_1$ and then compute the value
$v_2$. In the end, we transfer ${\bf v}$ back to the physical space
by ${\bf u}=R{\bf v}$.
\subsection{Time Evolution}
To evolve in time, we use the third-order TVD Runge-Kutta time
discretization (Shu \& Osher 1988). For system of ODEs $u_t=L(u)$,
the third order Runge-Kutta method is {\setlength\arraycolsep{2pt}
\begin{eqnarray*}
u^{(1)}&=&u^n+\Delta\tau L(u^n,\tau^n),\\
u^{(2)}&=&\frac{3}{4}u^n+\frac{1}{4}(u^{(1)}+\Delta\tau L(u^{(1)},\tau^n+\Delta\tau)),\\
u^{n+1}&=&\frac{1}{3}u^n+\frac{2}{3}(u^{(2)}+\Delta\tau
L(u^{(2)},\tau^n+\frac{1}{2}\Delta\tau)).
\end{eqnarray*}}


\begin{references}

\reference{} Adams, T.F. 1972, \apj, 174, 439

\reference{} Adams, T.F.  1975, \apj, 201, 350

\reference{} Adams, T.F., Hummer, D.G. \& Rybicki, G.B. 1971, J.
Quant. Spectrosc. Radiat. Transfer\apj, 11, 1365

\reference{} Avery, L.W., \& House, L.L. 1968, \apj, 152, 493

\reference{} Blanc, G. A, et al. 2010, arXiv:1011.0430

\reference{} Bonilha, J. R. M., Ferch, R., Salpeter, E. E., Slater, G., Noerdlinger, P. D. 1979,
\apj 233, 649

\reference{} Cockburn, B., Johnson, C., Shu, C-W. \& Tadmor, E. 1998,
Lecture Notes in Mathematics, 1697, 450

\reference{} Dijkstra, M., Haiman, Z. \& Spaans, M. 2006, \apj,
649, 14

\reference{} Dijkstra, M. \& Loeb, A.  2009, \mnras 400, 1109

\reference{} Draine, B.T. 2003, \apj, 598, 1017

\reference{} Draine, B. T.; Lee, H. M. 1984, \apj, 285, 89

\reference{} Fang, L.Z., 2009, Inter. J. Mod. Phys. D18, 1943

\reference{} Fardal et al. 2001,  \apj, 526, 505

\reference{} Field, G.B., 1958, Proc. IRE, 46, 240

\reference{} Field, G.B. 1959, \apj, 129, 551

\reference{} Haiman Z. et al. 2000, \apj, 537, L5

\reference{} Haiman Z. \& Spaans, M.  1999, \apj, 518, 138

\reference{} Hansen, M., \& Oh, S. P. 2006, \mnras, 367, 979

\reference{} Harrington, J.P. 1973, \mnras, 162, 43

\reference{} Hayes, M., et al. 2010, Nature, 464, 562

\reference{} Hayes, M., Schaerer, D., Östlin, G., Mas-Hesse, J. M., Atek, H. \& Kunth, D.
   2011, \apj, 730, 8

\reference{} Henyey, L. G. \& Greestein, J. L. 1941, \aj, 93, 70

\reference{} Hummer, D.G. 1962, \mnras, 125, 21

\reference{} Hummer, D.G. 1965, Mem. R. astr. Soc., 70, 1

\reference{} Hummer, D.G. 1969, \mnras, 145, 95

\reference{} Hummer, D.G. \& Kunasz, P.B. 1980, \apj, 236, 609

\reference{} Jiang, G. \& Shu, C.-W. 1996, J. Comput. Phys., 126, 202

\reference{} Latif, M.  et al. 2011, \mnras 413, L33

\reference{} Laursen, P., Sommer-Larsen, J., \& Andersen, A., 2009, \apj, 704, 1640

\reference{} Li, A., \& Draine, B. T. 2001, \apjl, 550, 213

\reference{} Liu, et al. 2007, \apj, 663, 1

\reference{} Neufeld, D. A.  1990, \apj, 350, 216

\reference{} Neufeld, D. A. 1991, \apj, 370, L85

\reference{} Osterbrock, D.E. 1962, \apj, 135, 195

\reference{} Pei, Y. C. 1992, \apj, 395, 130

\reference{} Qiu, J.-M., Feng, L.-L., Shu, C.-W. \& Fang, L.-Z. 2006, New Astronomy,
             12, 1

\reference{} Qiu, J.-M., Feng, L.-L., Shu, C.-W. \& Fang, L.-Z. 2007, New Astronomy,
             12, 398

\reference{} Qiu, J.-M., Shu, C.-W., Liu, J.-R. \& Fang, L.-Z. 2008, New Astronomy,
             13, 1

\reference{} Rybicki, G.B. 2006, \apj, 674, 709

\reference{} Rybicki, G.B. \& Dell'Antonio, I.P. 1994, \apj, 427, 603

\reference{} Rybicki G.B. \& Lightman, 1979 Radiative Processes in Astrophysics, (J. Wiley New York.

\reference{} Roy, I., Qiu J.-M., Shu C.-W. \& Fang L.-Z., 2009a New Astronomy 14, 513

\reference{} Roy, I. Xu, W., Qiu J.-M., Shu C.-W. \& Fang L.-Z., 2009b \apj, 694, 1121

\reference{} Roy, I. Xu, W., Qiu J.-M., Shu C.-W. \& Fang L.-Z., 2009c \apj, 703, 1992

\reference{} Roy, I, Shu, C.-W. \& Fang, L. Z. 2010, \apj, 716, 604

\reference{} Shu, C.-W. \& Osher, S., 1988, J. Comp. Phys., 77, 439

\reference{} Spaans, M. \& Silk, J., 2006, \apj, 652, 902

\reference{} Spitzer, L. \& Greenstein, J.L. 1951, \apj, 114, 407

\reference{} Stratta, G., Maiolino, R., Fiore, F. \&  D'Elia, V., 2007, 661, 9

\reference{} Unno, W. 1955, Publ. Astron. Soc. Japan.  7, 81

\reference{} Verhamme, A., Schaerer, D. \& Maselli, A. 2006, AA, 460, 397

\reference{} Verhamme, A.; Schaerer, D., Atek, H. \& Tapken, C. 2008, AA, 491, 89

\reference{} Weingartner, J. C.; Draine, B. T. 2001, \apj, 548, 296

\reference{} Wouthuysen, S. A. 1952, \aj, 57, 31

\reference{} Xu, W. \& Wu, X.-P. 2010, \apj, 710, 1432

\end{references}
\end{document}